\documentclass[twocolumn,tighten]{aastex62}

\usepackage{gensymb}
\usepackage{amsmath}
\usepackage{color}

\def\be{\begin{equation}}
\def\en{\end{equation}}

\def\H2O{H$_2$O}
\def\CO2{CO$_2$}
\def\DCE45{DCE$_{J-[4.5]}$}

\received{January 07, 2026}
\revised{March 28, 2026}
\accepted{June 23, 2026}
\submitjournal{ApJ}

\shortauthors{Manzo-Mart{\'i}nez et al.}

\begin{document}

\title{MODELING THE CURVED DUST SUBLIMATION FRONT IN PROTOPLANETARY DISKS: A POTENTIAL PROBE OF MIDPLANE TURBULENCE}

\author[0000-0001-6647-862X]{Ezequiel Manzo-Mart\'inez}
\affiliation{Instituto de Astronom\'ia, Universidad Nacional Aut\'onoma de M\'exico, Km. 107 Carretera Tijuana-Ensenada, 22860, Ensenada, Baja California, M\'exico} \email{emanzo@umich.edu}
\affiliation{Department of Astronomy, University of Michigan, 323 West Hall, 1085 South University Avenue, Ann Arbor, MI 48109, USA}
\author[0000-0002-1650-3740]{Ramiro Franco Hern\'andez}
\affiliation{Departamento de Física, CUCEI, Universidad de Guadalajara, Boulevard Marcelino García Barragán 1421, Olímpica, Guadalajara 44430, Jalisco, México}
\author[0000-0002-3950-5386]{Nuria Calvet}
\affiliation{Department of Astronomy, University of Michigan, 323 West Hall, 1085 South University Avenue, Ann Arbor, MI 48109, USA}
\author[0000-0001-9797-5661]{Jes\'us Hern\'andez}
\affiliation{Instituto de Astronom\'ia, Universidad Nacional Aut\'onoma de M\'exico, Km. 107 Carretera Tijuana-Ensenada, 22860, Ensenada, Baja California, M\'exico}
\author{Paola D'Alessio}
\affiliation{Instituto de Radioastronom\'ia y Astrof\'isica, Universidad Nacional Aut\'onoma de M\'exico, Apartado Postal 3-72, C.P. 58089 Morelia, Michoac\'an, M\'exico}
\author[0000-0002-7179-6427]{Rosa M. Torres}
\affiliation{Departamento de Física, CUCEI, Universidad de Guadalajara, Boulevard Marcelino García Barragán 1421, Olímpica, Guadalajara 44430, Jalisco, México}

\begin{abstract}

We present a new approach to calculate the geometry and emission of the dust inner wall in disks around T Tauri stars. This calculation follows a self-consistent approach given the disk structure and adopts a density-dependent sublimation temperature for the dust. We built spectral energy distributions (SEDs) of disk models with curved walls around a $0.5 \, M_{\odot}$ star, finding that  the curved wall starts at a radius of $\sim 0.11$ au and extends to $\sim 0.38$ au. The dependence on mass accretion rate, dust settling, and disk inclination on the resulting SEDs is explored, as well as the impact of the height of the midplane layer containing large millimeter-sized grains. To test our models, we compare synthetic near-IR colors from a grid of disk models with observed colors for a large sample of disk-bearing T Tauri stars located in Taurus, IC 348, and the Orion complex.  Most of the observed colors can be explained by combinations of mass accretion rates, dust settling, and inclinations within the expected ranges for T Tauri stars. However, populating the regions where observed JHK colors are most concentrated, requires the millimeter-size grains be spread up to $0.5-3$ scale heights above the midplane. This result contradicts expectations of rapid dust settling and suggests a high degree of turbulence capable of lifting large grains toward the upper disk layers. These findings provide insight into the dynamical conditions of the disk midplane near the star.

\end{abstract}


\section{Introduction}
\label{sec_intro}

Protoplanetary disks form as a consequence of the star formation process and are the natural sites where planet formation occurs. Studying the planet-formation process requires detailed knowledge of the physical properties of the innermost disk regions at various evolutionary stages. This is essential not only because planets characteristics may be determined by disk properties during formation, but also because such studies can help us understand the origin and properties of the Solar System.\\

Protoplanetary disks have a wide range of temperatures, and close to the star, the high temperatures sublimate the dust, creating a sublimation front commonly called the wall or the inner rim \citep{Dullemond_2010}. This is the region where dust evaporates, and thus the wall surface is directly irradiated by the star. The emission from the irradiated inner rim largely dominates the SED at near-IR wavelengths. Moreover, the near- and mid-IR fluxes, as well as the shape and intensity of the silicate feature at 10 $\mu$m, characteristic of Classical T Tauri Stars (CTTSs), are direct tracers of the properties and composition of the dust in the wall. The inner disk regions typically evolve over timescales of 5–10 Myr \citep{Hernandez_2007a, Briceno_2019}, with mid-infrared SED slopes decreasing as stellar populations age, reflecting the evolutionary stage of the inner disk  \citep{Hernandez_2007b}. Improving our understanding of the innermost disk regions is therefore critical for deciphering the processes that drive this evolution.\\

Several studies have addressed this by either fitting SEDs using disk models where the observed excess over the photosphere originates from the disk dust component \citep[e.g.][]{McClure_2013, Ribas_2020, Rilinger_2023}, or by using numerical simulations that incorporate multiple physical processes acting simultaneously \citep[e.g.][]{Birnstiel_2011, Birnstiel_2015, Gole_2020, Cevallos_2025, Mori_2025, Vaikundaraman_2025}. These approaches have complemented each other, providing valuable insights into disk structure and the physical mechanisms that drive disk evolution. Curved inner rims in disks around Herbig Ae stars were studied by \citet{Flock_2016, Flock_2017} using hydrodynamical models including radiation, while \citet{Flock_2019} investigated the formation and migration of super-Earth and Earth-like planets near the location of the wall. Numerical studies by  \citet{Ueda_2017, Ueda_2019} were used to derive an analytical expression for different zones in the inner rim, and to study the effects of dust accumulation in the disk inner regions. More recently, a multi-band study of the inner rim of HD 163296 was performed by \citet{Chrenko_2024}, which points out the challenges in modeling the wall emission. Moreover, an analysis of the changes produced in the wall geometry of magnetized disks when different dust species are considered was made by \citet{Flock_2025}. \\

Over the past decades, substantial improvements in the study of the innermost disk regions have been achieved due to the advent of near-IR photometry and spectroscopy from facilities such as {\it Spitzer} (IRS and IRAC), and the Wide-field Infrared Survey Explorer (WISE) \citep{Hernandez_2007b, Furlan_2011, Manzo_2020}. More recently, JWST mid-infrared instrument (MIRI) spectra have enabled more detailed studies of these regions. Early studies modeled the inner rim emission assuming a vertical wall with a single dust sublimation temperature \citep[e.g.][]{Natta_2001, Dalessio_2005}. It was later proposed that the wall is curved as a result of vertical density and temperature gradients in the disk, as shown by several studies \citep{Isella_2005, Tannirkulam_2007, Kama_2009, Dullemond_2010}. Following these results, \citet{McClure_2013} used a two-layered wall model to mimic the expected curvature. This model successfully reproduced the SEDs of Taurus sources with high IR excesses that could not be explained by a single vertical wall. \\
 
Although direct imaging of the wall in CTTSs remains elusive, recent JWST data and advances in near-IR observations and interferometry highlight the need for more physically realistic models of the inner rim. In the last decade, high angular-resolution observational techniques, such as near-infrared VLTI interferometry, have significantly improved our ability to image and constrain the location of the dust inner rims for intermediate-mass Herbig Ae/Be stars \citep[e.g.][]{Anthonioz_2015, Ibrahim_2023} and have begun to make progress toward imaging the wall in the CTTS regime \citep{Gravity_2021}. Similarly, PIONIER H-band continuum observations have probed the emission within a few astronomical units (au) of these stars \citep{Lazareff_2017, Kluska_2020}. 
The ALMA sub-mm interferometer is currently unable to resolve the inner rim of TTSs, i.e., scales of $\sim 0.1-0.5$ au, even for nearby systems. Given these observational limitations, SED modeling therefore remains one of the most effective tools for studying the wall properties, emphasizing the need for improved wall models. Moreover, recent studies have proposed that the location of the wall may set the location of the inner orbit of small planets, while the magnetospheric radius could set the inner orbit for hot Jupiters \citep{Sun_2025, Mendigutia_2024}. A detailed model of the wall allows us both to reproduce observed SEDs and to generate wavelength-dependent brightness distributions, from which interferometric visibilities can be computed and compared with observations.\\

Motivated by these observational limitations, together with the advances in infrared and interferometric observations, we developed the curved wall model presented in this work. The structure of this paper is as follows. In Section \ref{sec_observations}, we describe the CTTS samples used to perform tests of the model. Section \ref{sec_model} presents the model for the curved wall and its main assumptions. The resulting wall geometry and emission are discussed in Section \ref{sec_results}. In Section \ref{sec_discussion}, we analyze the implications of our results, and finally, Section \ref{sec_summary} summarizes and
presents the main conclusions.\\

\section{Observational sample}
\label{sec_observations}

To test our curved wall models, we selected a sample of CTTSs spanning a wide age range from 1 to 11 Myr, and spectral types from K0 to M6. The sample includes sources located in Taurus, IC 348, and the Orion star-forming complex. Below, we describe the criteria used for the sample selection.\\

\subsection{Taurus}

The Taurus complex consists of several molecular clouds located at distances of 130 to 200 pc \citep{Galli_2019} and with an estimated age of 1--3 Myr \citep{Luhman2023} making it an ideal region to study disks at early evolutionary stages. Our sample consists of 138 stars from \citet{Esplin_2019} classified as full disks, i.e., disks without holes or gaps that would show a deficit at near-IR wavelengths in the SEDs \citep{Espaillat_2014}. The JHK photometry, spectral types, and the extinction values in the J band ($A_{\rm J}$) were taken from \citet{Esplin_2019}.

\subsection{IC 348}

IC 348 is a young cluster \citep[$\sim 2-3$ Myr][]{Herbst_2008}, located at a distance of 320 pc \citep{Ortiz_2018}. For this cluster, we used the  sample from \citet{Luhman_2016}, which reports spectral types and $A_{\rm J}$. We selected disk-bearing stars classified as sources with infrared excess in that work, based on mid-IR photometry from the {\it Spitzer Space Telescope}, reported mainly by \citet{Lada_2006}. Using this photometry, we obtained a sample of stars with full disks by selecting stars with $\alpha_{\rm IRAC}$ values in the range -1.8 to 0 \citep{Lada_2006}, where $\alpha_{\rm IRAC}$ is the spectral slope between 3.6 and 8 \micron. We also included stars with full disks from \citet{Lada_2006} that were not studied by \citet{Luhman_2016}, applying the same $\alpha_{\rm IRAC}$ criterion to select sources with full disks, with spectral types and visual extinction ($A_{\rm V}$) reported in that work. The final IC 348 sample consists of 142 stars.

\subsection{The Orion star-forming complex}

The Orion star-forming complex consists of several stellar populations spanning ages of approximately 1--11 Myr, 
located at distances $\sim 350- 400$ pc \citep{Briceno_2019, Kounkel_2018}. For this region, we used a sub-sample of stars from \citet{Briceno_2019} located in the Orion OB1 association
classified as CTTS and with spectral types and $A_{\rm V}$ reported in that work. The final Orion sample consists of 208 stars.

\subsection{JHK colors and correction for interstellar reddening}
\label{subsec_colors}

\begin{figure}
    \centering     \includegraphics[width=1\linewidth]{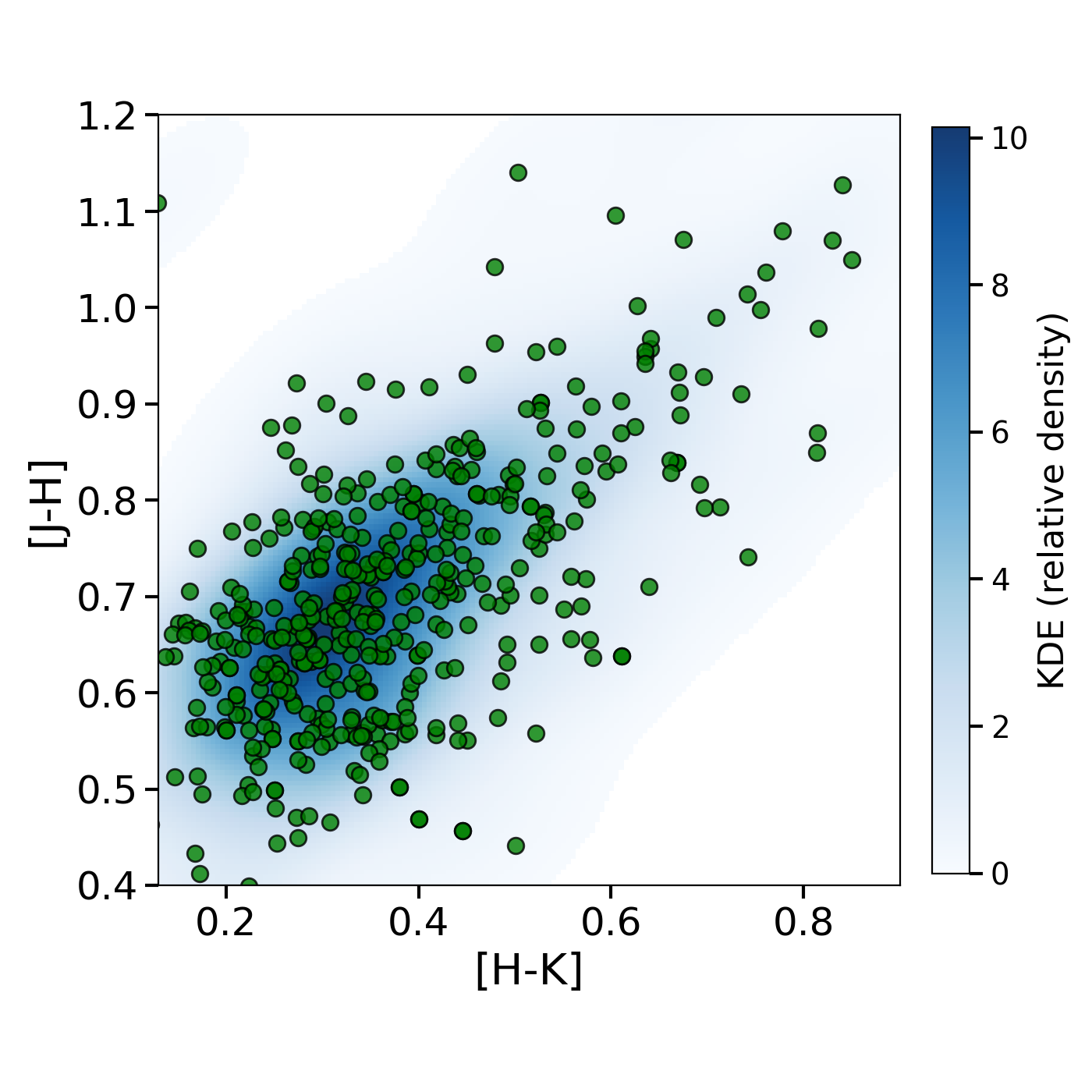}
    \caption{Color-color diagram in the JHK bands for our CTTSs sample (green points). The shaded blue region is the corresponding normalized two-dimensional KDE map, assuming a Gaussian kernel.}
    \label{fig:KDE_obs}
\end{figure}

We used 2MASS photometry \citep{2MASS_2006} to build the observed JHK color-color diagrams for our samples. Using the reported values of either $A_{\rm V}$ or $A_{\rm J}$, we corrected the ${\rm [J-H]}$ and ${\rm [H-K]}$ colors for interstellar reddening. The extinction coefficients ($A_{\rm \lambda}/A_{\rm V}$) for the 2MASS J, H, and K bands were obtained from \citet{Cardelli_1989} with R$_{\rm V}=3.1$, assuming effective wavelengths for each band of 1.235, 1.662, and 2.159 \micron, respectively. In Figure \ref{fig:KDE_obs} we show the dereddened colors in the ${\rm [J-H]}$ vs ${\rm [H-K]}$ plane for the full CTTSs sample, and represent this distribution of colors using a two-dimensional kernel density estimate (KDE). We adopted a Gaussian kernel implemented via \texttt{scipy.stats.gaussian\_kde}, and used Scott's rule for bandwidth selection., i.e., a single global smoothing factor that scales the sample covariance matrix to define the kernel covariance. We evaluated the KDE on a uniform $300 \times 300$ grid spanning the full range of the observed colors, $x={\rm [H-K]}\in[\min(x),\max(x)]$ and $y={\rm [J-H]}\in[\min(y),\max(y)]$ for the CTTS sample. No additional weighting of individual sources was applied. The resulting KDE map (blue shaded region) is a probability density estimate normalized to integrate to unity over all $({\rm [J-H]}, {\rm [H-K]})$ values. For visualization we mask extremely low-density values ($<10^{-8}$), and display the map with a linear colormap. For reference, we overplotted as green points the individual CTTS colors used to compute the KDE.\\

\section{Model for the curved wall}
\label{sec_model}

\subsection{The D'Alessio models}
\label{subsec_Dalessio_models}

The D'Alessio Irradiated Accretion Disk (DIAD) models \citep{Dalessio_1998, Dalessio_1999, Dalessio_2001, Dalessio_2005,  Dalessio_2006} self-consistently calculate
the vertical and radial structure of protoplanetary disks heated by viscous dissipation as well as irradiation from the central star and the accretion shocks. The models solve a coupled set of equations for the disk structure in an iterative scheme to find the hydrostatic equilibrium and the transport of energy \citep{Dalessio_1998}. The main input parameters of the models include the stellar properties ($M_{*}$, $T_{\rm eff}$, $R_{*}$), the mass accretion rate $\dot{M}$ onto the star, the viscosity parameter $\alpha$ following the prescription of \citet{Shakura_1973}, the disk radius, and the distance to the system, which is required for computing observable quantities. The models include dust settling of large grains towards the midplane, quantified by the parameter $\epsilon=\zeta_{\rm small}/\zeta_{\rm std}$, defined as the dust-to-gas mass ratio in the disk upper layers relative to the typical interstellar value $\zeta_{\rm std}=0.01$ \citep[see][for more details]{Dalessio_2006}. The dust grain populations follow a size distribution with maximum sizes of $a_{\rm max,small}=0.25~\mu$m for the atmospheric grains, and
$a_{\rm max,big}=1$~mm for the midplane layer. The large and small grains are separated by the transition height $z_{\rm big}$, defined as the vertical extent of the large-grain layer above the midplane, expressed in units of the local gas scale height $H$. Here $H=c_{s}/\Omega_{k}$, where $c_{s}=(kT/\mu m_{\rm H})^{1/2}$ is the local sound speed evaluated at the midplane temperature, $\mu$ the mean molecular weight,
and $\Omega_{k}=(GM_{*}/R^{3})^{1/2}$  the Keplerian angular velocity. We use these models as input for our curved wall calculations. The models were used to construct a grid with the parameter ranges described in Table \ref{tab:parameters}, which span a comprehensive range of observed mass accretion rates, inclinations from close to face-on ($\cos(i)=0.9)$ to nearly edge-on ($\cos(i)=0.1$) configurations, $\epsilon$ values that correspond to no settling ($\epsilon=1$) to highly settled ($\epsilon=0.0001$) disks, and $z_{big}$ values from very close to the midplane (0.1 H) up to 3 gas scale heights. The parameter space was discretized by sampling $\cos(i)$ in steps of 0.1, $\dot{M}\in\{1\times 10^{-7}, 5\times 10^{-8}, 1\times 10^{-8},5\times10^{-9},1\times10^{-9},5\times10^{-10}\}\,M_{\odot}\,{\rm yr}^{-1}$, dust settling $\epsilon\in\{1, 0.1,0.01,0.001,0.0001\}$, and $z_{\rm big}$ in steps of 0.5. For the central star we considered $0.3 \, M_{\odot}$ and $0.5 \,M_{\odot}$. We used this grid to compare our models with observations (see Section \ref{subsec_comparison}).

\begin{table}
\centering
\caption{Range of values used in the model grid}
\begin{tabular}{l l l}
\hline
Parameter & min  & max \\
\hline
\hline
$\epsilon$ & 0.0001 & 1 \\
$\dot{M} (M_{\odot}\,\mathrm{yr^{-1}})$ & $5\times10^{-10}$ & $1\times10^{-7}$ \\
$\cos(i)$ & 0.1 & 0.9 \\
$z_{\rm big} ({\rm H})$  &  0.1 & 3\\
\hline
\hline
\end{tabular}
\label{tab:parameters}
\end{table}

\subsection{The geometry of the wall}

\begin{figure}[ht]
    \centering
\includegraphics[width=0.83\linewidth]{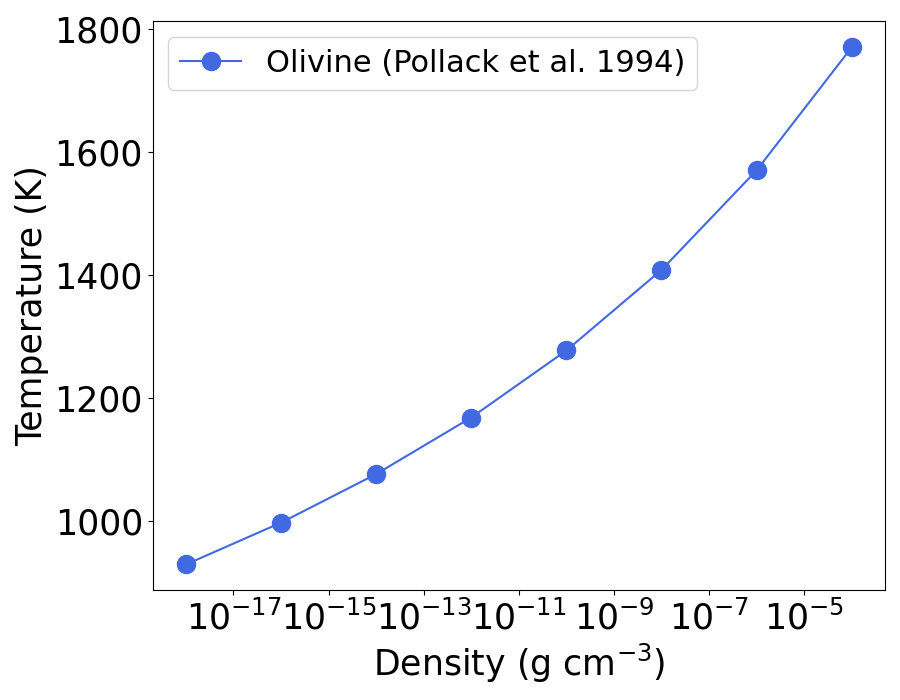}
    \caption{Sublimation temperature vs density for olivine from \citet{Pollack_1994}.}
    \label{fig_temp}
\end{figure}

To define the geometry of the curved wall, we use the structure of a given disk model, namely the vertical and radial distributions of temperature $T(R,z)$, density $\rho(R,z)$, and pressure $P(R,z)$. The geometry is obtained using equation \ref{eq_rsub} from \citet{Dalessio_2004}:

\begin{equation}
    R_{sub,i,j}=\left[\left(\frac{L_{*} + L_{shock}}{16\pi \sigma_{R}}\right)\left(2+\frac{\bar{\chi_{*}}_{i,j}}{\bar{\kappa_{d}}_{i,j}}\right)\right]^{1/2}\frac{1}{T_{\rm sub}(\rho_{i,j})^{2}}
    \label{eq_rsub}
\end{equation}

\noindent
which is applied to search for all radii $R_{i}$, and at each height $z_{j}$, the points $(R_{i}, z_{j})$ in the disk that match  $R_{sub,i,j}$. The temperature  $T_{\rm sub}(\rho_{i,j})$ is the density-dependent sublimation temperature for a given dust species. We assumed that the dust was made of amorphous olivine and adopted $T_{\rm sub}(\rho)$ from \citet{Pollack_1994}, which is shown in Figure \ref{fig_temp}. In equation (\ref{eq_rsub}), $L_{*}$ denotes the stellar luminosity, while $L_{\rm shock}$ is the luminosity produced by accretion shocks, which we take as $L_{\rm shock}=0.8 \, L_{\rm acc}$, with
$L_{\rm acc}=GM_{*}\dot{M}/R_{*}$
following \citet{Calvet_1998}. Here, $\sigma_{R}$ is the Stefan-Boltzmann constant. The total mean opacity at the stellar wavelength range, $\bar{\chi_{*}}_{i,j}$, is the sum of the opacities at the stellar wavelength range of the small and large grains, weighted by their respective abundances. In turn, the mean opacities at the stellar range are given by
$\bar{\chi_{*}}=\int_{0}^{\infty}\chi(\lambda)B_{\nu}(T_{*})d\lambda/\int_{0}^{\infty}B_{\nu}(T_{*})d\lambda$, and are calculated with the opacity $\chi(\lambda)$ corresponding to each type of grains. 
Similarly, the mean true opacity at the disk wavelength range,
$\bar{\kappa_{d}}_{i,j}$,
is the sum of the true mean opacities of small and large grains, weighted by their respective abundances, where the true mean opacity is given by $\bar{\kappa_{d}}=\int_{0}^{\infty}\kappa(\lambda)B_{\nu}(T_{sub}(z))d\lambda/\int_{0}^{\infty}B_{\nu}(T_{sub}(z))d\lambda$ \citep{Dalessio_2004}. It must be noticed that there is no single value for the mean opacities $\bar{\chi_{*}}$ and $\bar{\kappa_{d}}$ in the wall geometry calculation; instead they are computed self-consistently given the specific set of parameters of the model, i.e., dust composition, grain sizes, and abundances at each point $(R_{i},z_{j})$. Specific values for the opacities of our fiducial model are discussed later in Section \ref{subsec_wall_results}.

\subsection{Wall emission}
\label{subsec_the_emission}

\begin{figure}[ht]
    \centering
    \includegraphics[scale=0.90]{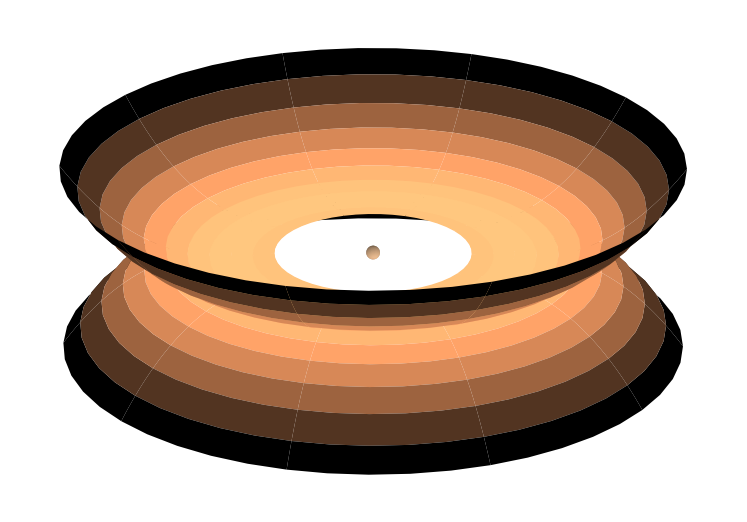}
    \caption{Visualization of the shape of the curved wall model as seen at an inclination $\cos(i)=0.4$. The parameters are $\dot{M}=1\times10^{-8} \, M_{\odot} \, {\rm yr}^{-1}$, $\epsilon=0.01$, and $z_{\rm big}=0.1$ H, for a $0.5 \, M_{\odot}$ star. Note that the lower wall is not visible except through the central hole. The star is shown at the center.}
    \label{fig_conical}
\end{figure}

To compute the emission from the curved wall, we first determined the angle between the incident stellar radiation and the normal vector of each surface element of the wall. For this, we fitted a third-degree polynomial to the curved wall geometry (see Section \ref{subsec_SEDs_curved_walls}) and discretized the surface area into a grid of 100$\times$100 equally sized surface elements in polar coordinates.  For the surface elements that are not hidden behind the disk or other parts of the wall, we integrated the radiative transfer equation along the line of sight to compute the total emission. The inclination of the system is given by the angle $i$ between the rotation axis of the system and the line of sight. The wall is not hidden by the star since it is located at $\sim 0.11$ au, or $\sim 11\,R_{*}$, (see Section \ref{subsec_wall_results}). Figure \ref{fig_conical} shows a scaled example of the geometry and appearance of a typical curved wall produced by our model. 
Once the wall emission is computed, we can build the total SED of a model, which includes the contributions from different components: the stellar photosphere, the disk, the curved wall, and the accretion shocks.\\ 

\begin{figure*}[ht]
    \centering
    \includegraphics[scale=0.31]{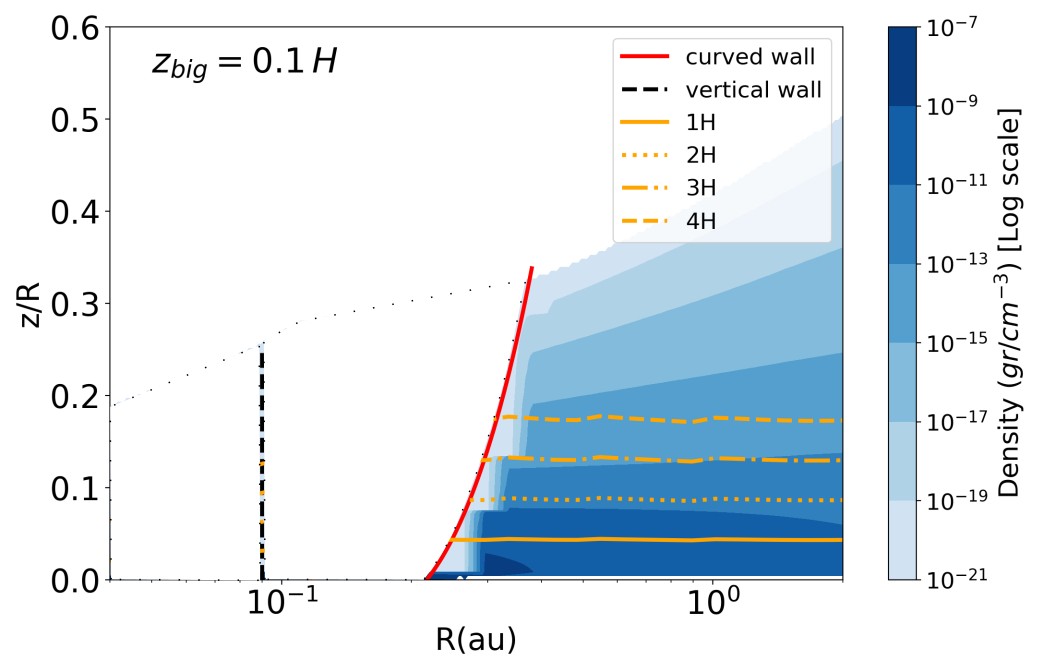}
    \includegraphics[scale=0.31]{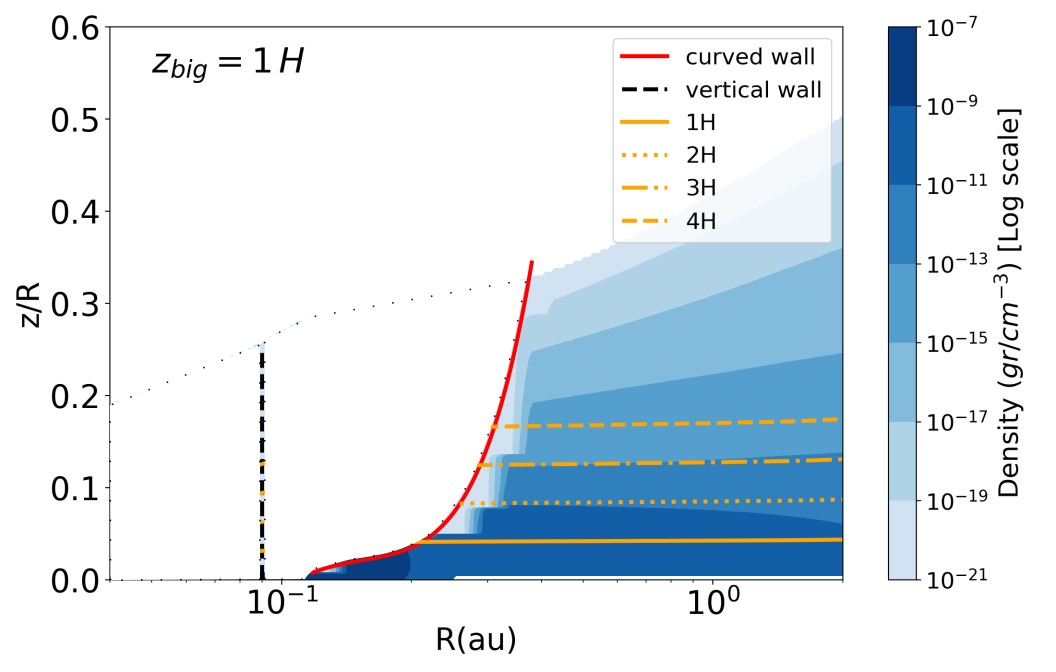}
    \includegraphics[scale=0.31]{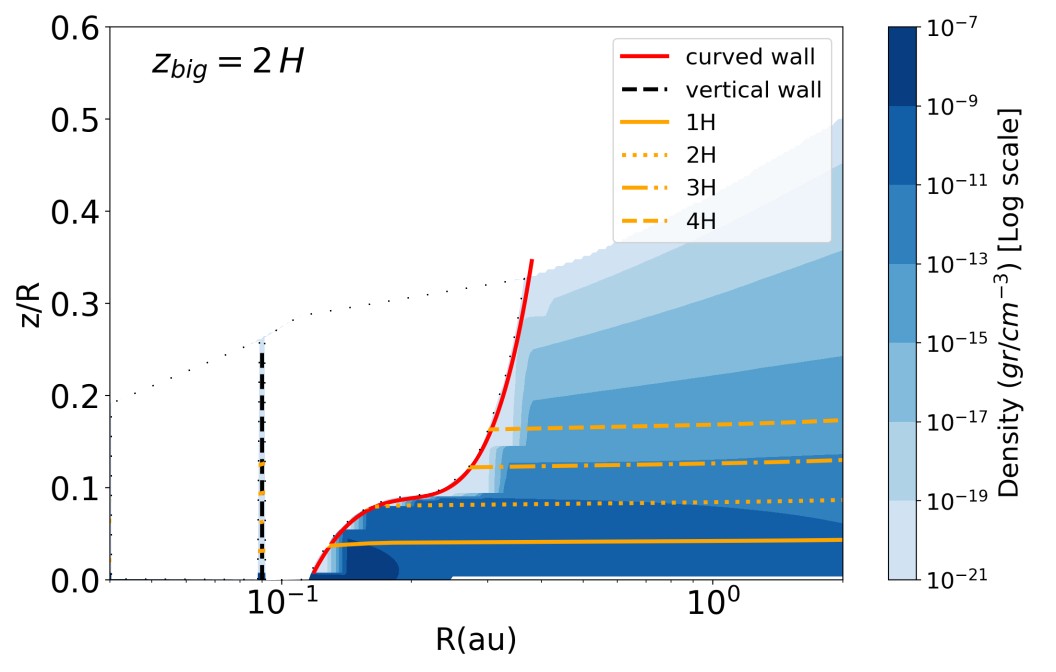}
    \includegraphics[scale=0.31]{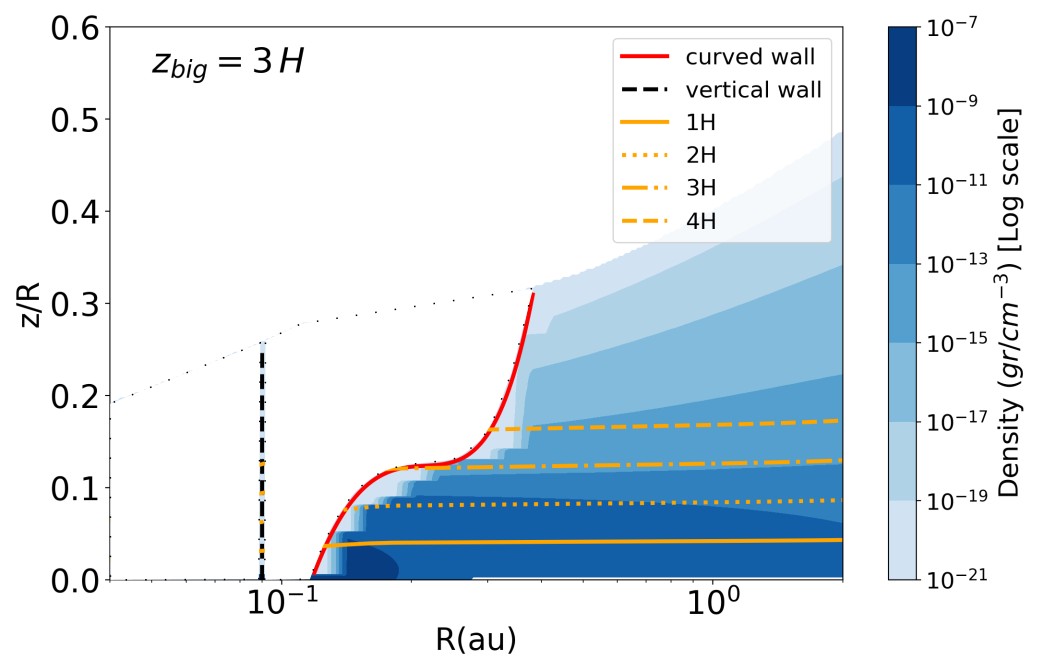}
    \caption{Third-degree polynomial fit to the geometry of the wall (red line) shown in the disk dust density plot (blue region), for various $z_{\rm big}$ values. The disk parameters are $\dot{M}=1\times10^{-8} \, M_{\odot} \, {\rm yr}^{-1}$ and $\epsilon=0.01$, for a $0.5 \, M_{\odot}$, 1 Myr old star. The thick black dashed line is the location of a vertical wall with $T_{\rm sub}=1400$ K, while in orange we show the scale height at different values, as indicated by the labels.}
    \label{fig_disks_density_wall}
\end{figure*}

\section{Results}
 \label{sec_results}

\subsection{Curved walls for various disk models}
\label{subsec_wall_results}

Figure \ref{fig_disks_density_wall} shows the disk density structure and the corresponding polynomial fit to the curved wall geometry (red solid line), for a reference model with $\dot{M}=1\times10^{-8} \, M_{\odot} \, {\rm yr}^{-1}$, $\epsilon=0.01$, and a $0.5 \, M_{\odot}$, 1 Myr old star, with $T_{\rm eff}=3769$ K and $R_{*}= 2.06 \, R_{\odot}$. Each panel corresponds to a different value of $z_{\rm big}$. The resulting disk density is shown in blue, and the material located interior to the wall (to the left of the red line) was removed after defining the curved wall geometry. In Figure \ref{fig_disks_density_wall},  we see that the wall location and radial extent, ranging from $\sim 0.12-0.33$ au, are consistent with dust sublimation radius estimates for CTTSs \citep[e.g.][]{Akeson_2005, McClure_2013, Gravity_2021}. For the models explored here, Figure \ref{fig_disks_density_wall} shows that the radial extent of the wall does not vary significantly with $z_{\rm big}$, except for the model with $z_{\rm big}=0.1$, in which the wall starts at $\sim 0.2$ au. In this model, most of the large grains are confined to the midplane; therefore, the wall location is primarily determined by the small grains. 

The wall geometry is curved as a consequence of the vertical gradients in the disk density and temperature,
together with the density dependence of the sublimation temperature $T_{\rm sub}(\rho)$ used in equation \ref{eq_rsub}. In addition, the total mean opacities vary with $z_{\rm big}$, as they depend on the relative abundances of large and small grains at each height. Since millimeter-sized grains can survive at smaller radii than micron-sized grains, models with larger $z_{\rm big}$ values exhibit a more vertically extended layer of large grains, which is reflected in the inner rim geometry.

For comparison, Figure \ref{fig_disks_density_wall} also shows the location of the vertical wall (black dashed line) predicted by previous models with a single population of grains, assuming a fixed sublimation temperature of 1400 K and a maximum grain size of $a_{\rm max}=0.25 \, \mu$m.
In the vertical wall approximation, a single sublimation temperature and a fixed maximum grain size are assumed, whereas the curved wall model accounts for the density dependence of $T_{\rm sub}$ and for the variations of the relative abundances of small and large grains with height. While vertical wall models typically adopt $T_{\rm sub}=1400$ K, the curved wall calculation uses density-dependent sublimation temperatures taken from \citet{Pollack_1994}. For the midplane densities of the models shown in Figure \ref{fig_disks_density_wall}, the corresponding sublimation temperature is $T_{\rm sub}=1350$ K, and it decreases with height as the density decreases towards the upper layers, explaining  why the vertical wall with $T_{\rm sub}=1400$ K is located closer to the star than the curved wall.

\begin{figure*}
    \centering
    \includegraphics[scale=0.35]{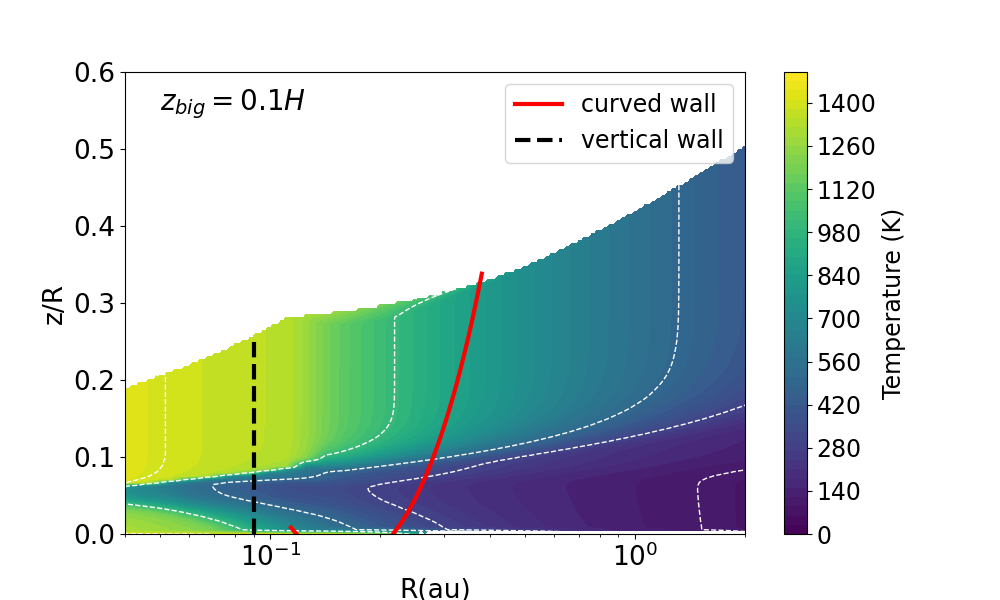}
    \includegraphics[scale=0.35]{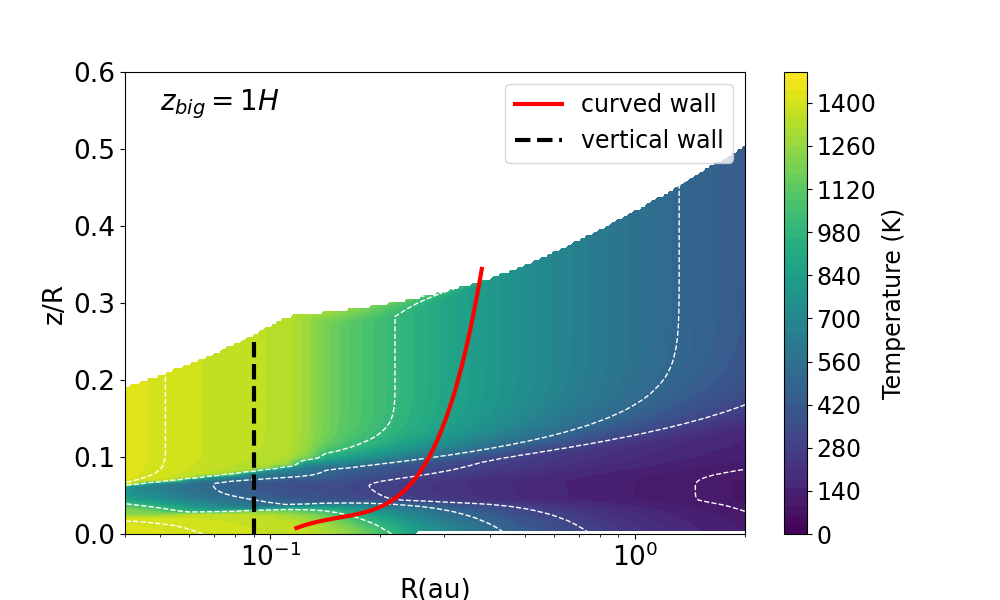}
    \includegraphics[scale=0.35]{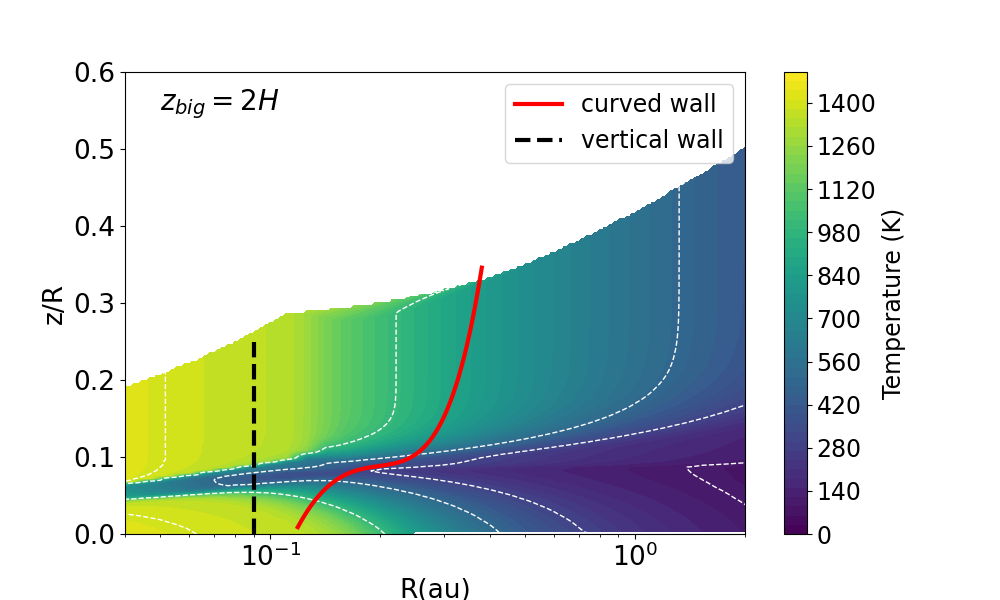}
    \includegraphics[scale=0.35]{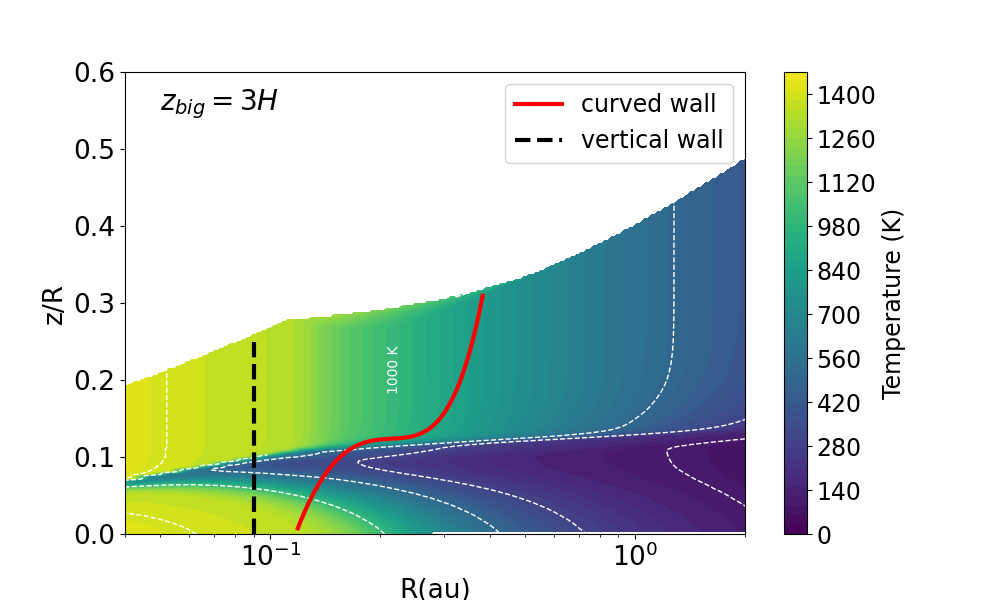}
    \caption{Third-degree polynomial fit to the geometry of the wall (red solid thick line) shown in the disk temperature structure, for various $z_{\rm big}$ values. The disk parameters are $\dot{M}=1\times10^{-8} \, M_{\odot} \, {\rm yr}^{-1}$ and $\epsilon=0.01$, for a $0.5 \, M_{\odot}$, 1 Myr old star. The black line is the location of a vertical wall with $T_{\rm sub}=1400$ K. Dashed lines are temperature isocontours for: 100, 300, 500, 1000, and 1400 K, from right to left.}
    \label{fig_disks_temp_wall}
\end{figure*}

Figure \ref{fig_disks_temp_wall} shows the disk temperature structure, with the polynomial fit to the curved wall indicated by the thick red line, and the vertical wall with the black line. Close to the star, the disk midplane temperature increases due to viscous heating at radii $\lesssim 0.3$ au for $z_{\rm big}=0.1$, and out to $\lesssim 0.8$ au for $z_{\rm big}=3$ H. In the upper layers, the disk is heated by irradiation from the central star. Farther from the star, viscous dissipation is no longer significant, and irradiation from the star dominates the disk heating. Figure \ref{fig_disks_temp_wall} includes temperature isocontours for 100, 300, 500, 1000, and 1400 K, shown as dashed white lines. Since the wall is frontally illuminated by the star and shock, it is hotter than the disk material immediately behind it for a given height.

\begin{figure*}
    \centering
    \includegraphics[scale=0.45]{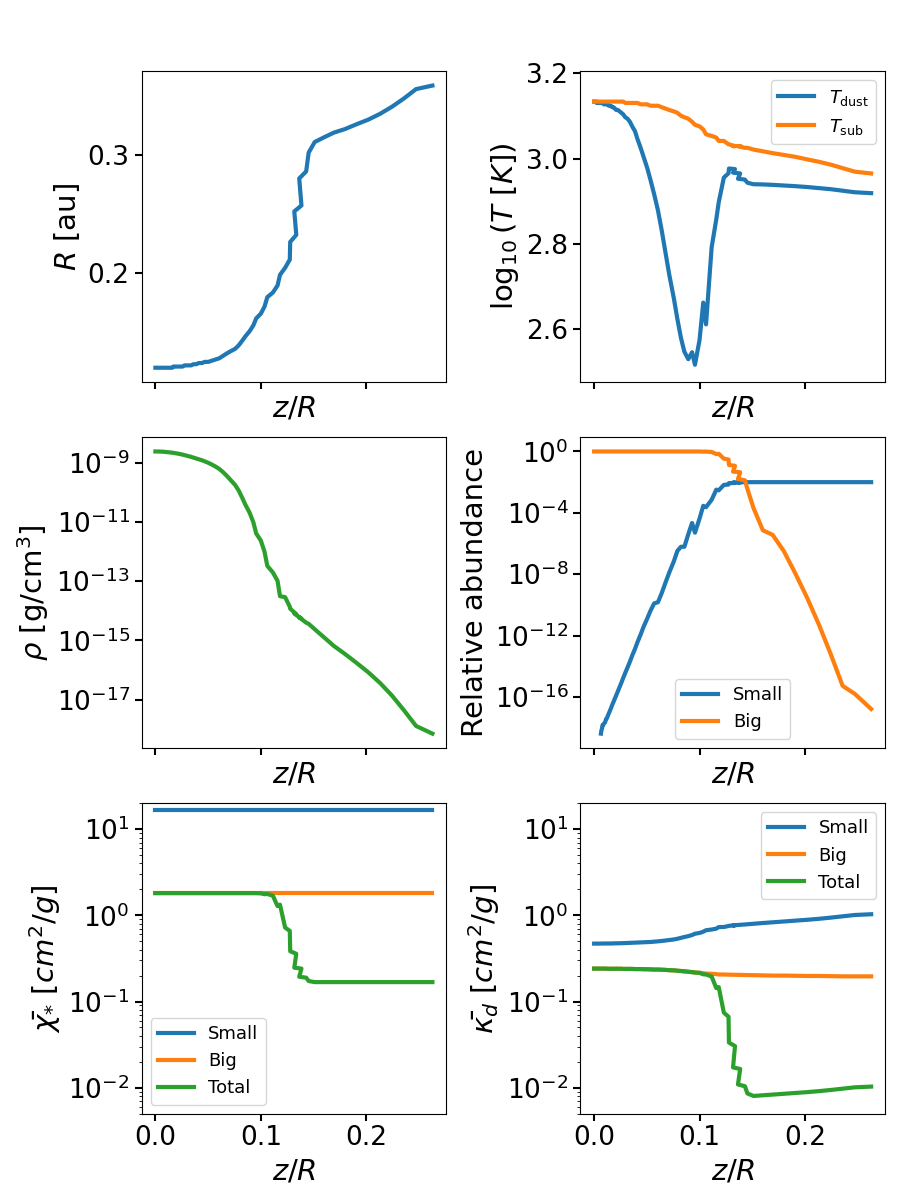}
    \caption{Physical properties at the wall as a function of the disk height, for a model with $z_{\rm big}=3 H$, $\dot{M}=1\times 10^{-8}M_{\odot} \, {\rm yr}^{-1}$, $\epsilon=0.01$, and a $0.5 \, M_{\odot}$, 1 Myr star. We show the wall geometry (upper left), the dust and sublimation temperatures (upper right),  the disk density (middle left), the abundances of large and small grains  (middle right),  the total mean opacity (lower left) and the mean true opacity (lower right). The total mean opacity and the true mean opacity (green lines in the lower panels) are obtained as the sum of the mean opacities of the small and large grain populations, each weighted by their respective dust-to-gas mass ratios.}
    \label{fig_wall_structure}
\end{figure*}

Figure \ref{fig_wall_structure} shows several properties along the curved inner rim as a function of disk height, for the model with $z_{\rm big}= 3$ H presented in Figures \ref{fig_disks_density_wall} and \ref{fig_disks_temp_wall}. In the first row, the left panel shows the wall geometry, while the right panel displays the sublimation temperature $T_{\rm sub}$, that is, the wall temperature, along with the dust temperature of the disk behind it. The wall temperature is higher, ranging 910 K to 1360 K, since it is frontally illuminated by the star, while the disk dust temperature is the result of heating by viscous dissipation in the midplane, and stellar irradiation in the disk the upper layers. This is why the curved wall (red solid line) in Figure \ref{fig_disks_temp_wall} intersects regions where the dust behind has relatively low temperatures ($\sim 500$ K). The disk temperature may actually be lower, since it may be shielded by the wall, but the emission from the wall would not be significantly affected. In the second row, the left panel shows the density, which decreases with height, explaining the decrease in $T_{\rm sub}$. The right panel displays the relative abundance of small and large grains as a function of height. The small grains dominate in the upper layers and are depleted by the $\epsilon$ value. The large grains extend up to $z \sim 0.15 \, R \sim 0.03 $ au above the midplane. Finally, the third row of Figure \ref{fig_wall_structure} shows the behavior of the mean opacities. 
In the left panel, we show the mean opacities in the stellar wavelength range for the small-grain (blue line) and large-grain (orange line) populations, together with the total mean opacity in the stellar wavelength range, $\bar{\chi_{*}}$  (green line). The total opacity is computed as the sum of the mean opacities of the small and large grain populations, each weighted by their respective dust-to-gas mass ratios. The right panel shows the same for the true mean opacity $\bar{\kappa_{d}}$. For this particular model $\bar{\chi_{*}}=1.68\times 10^{-1}$ and $\bar{\kappa_{d}}=1.06\times 10^{-2}  \, {\rm cm^{2} \, {\rm g^{-1}}}$ in the disk surface, while $\bar{\chi_{*}}= 1.81$ and  $\bar{\kappa_{d}}=0.24 \, {\rm cm^{2} \, {\rm g^{-1}}}$ in the midplane. The total opacities decrease as the height increases because of the depletion of small grains in the upper layers.

\subsection{The SEDs from curved walls}
\label{subsec_SEDs_curved_walls}

\begin{figure}
    \centering
    \includegraphics[scale=0.35]{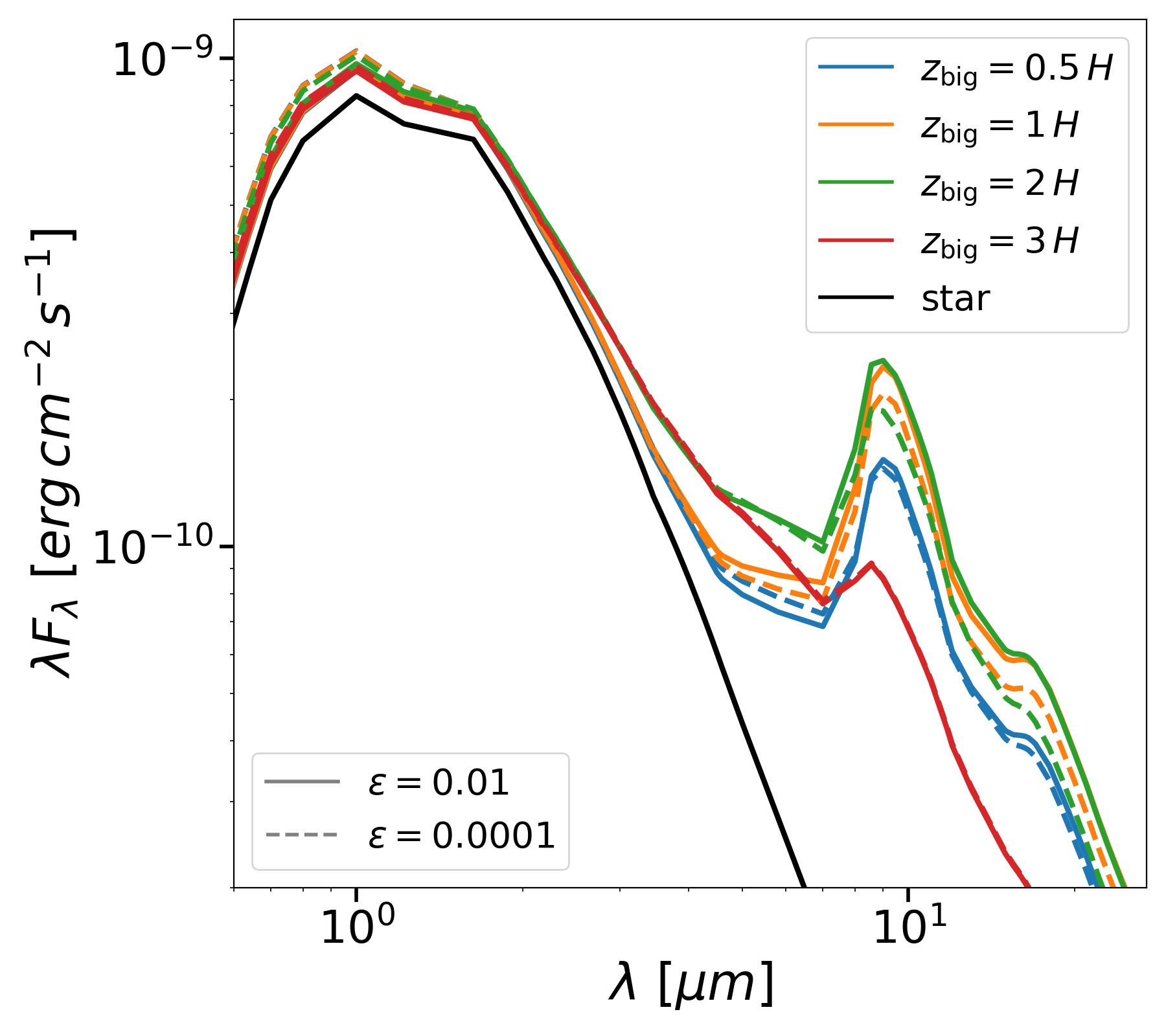}
    \caption{SEDs of the curved wall plus star for various $z_{\rm big}$ values: 0.5 H (blue), 1 H (orange), 2 H (green), and 3 H (red). The stellar photosphere is shown in black. The disk model has $M_{\ast}=0.5 \, M_{\odot}$ and $\dot{M}=1 \times 10 ^{-8} M_{\odot} \, {\rm yr}^{-1}$, with $\cos(i)=0.5$, for $\epsilon=0.01$ (solid) and $\epsilon=0.0001$ (dashed).}
    \label{fig_walls_SEDS}
\end{figure}

The emergent flux of the curved wall and the photosphere is shown in Figure \ref{fig_walls_SEDS} for $z_{\rm big}=0.5$, 1, 2, and 3 H, for disk models with $\epsilon=0.01$ (solid line) and $\epsilon=0.0001$ (dashed line). The stellar photosphere
 is the solid black line. The wall emission increases with increasing $z_{\rm big}$ at wavelengths $\lambda \gtrsim 1.7 \, \mu$m.
 For $z_{\rm big} = 3$ H, the effect is the opposite for $\lambda \gtrsim 4.5,\mu$m. In this case, the increased abundance of millimeter-sized grains in the upper layers reduces the strength and flattens the shape of the $10 \, \mu {\rm m}$ silicate feature, which is produced by small, hot grains in the disk atmosphere \citep[see][]{Dalessio_2006}. The effect of $\epsilon$ is negligible for small values of $z_{\rm big}$, but becomes more important for larger $z_{\rm big}$, at wavelengths $\lambda > 10 \, \mu$m.

As discussed above, the geometry of the curved wall is determined by the density-dependent sublimation temperature $T_{\rm sub}$. As a result, the curved wall emission differs from that of
a vertical wall, which assumes a single sublimation temperature.
Another key difference concerns the viewing geometry: for disks viewed complete face-on, a vertical wall does not contribute to the emission since its surface is parallel to the line of sight, whereas the curved wall reaches its maximum emission for face-on systems, in which the entire wall surface is visible as a bright ring.

\begin{figure*}
    \centering
    \includegraphics[scale=0.5]{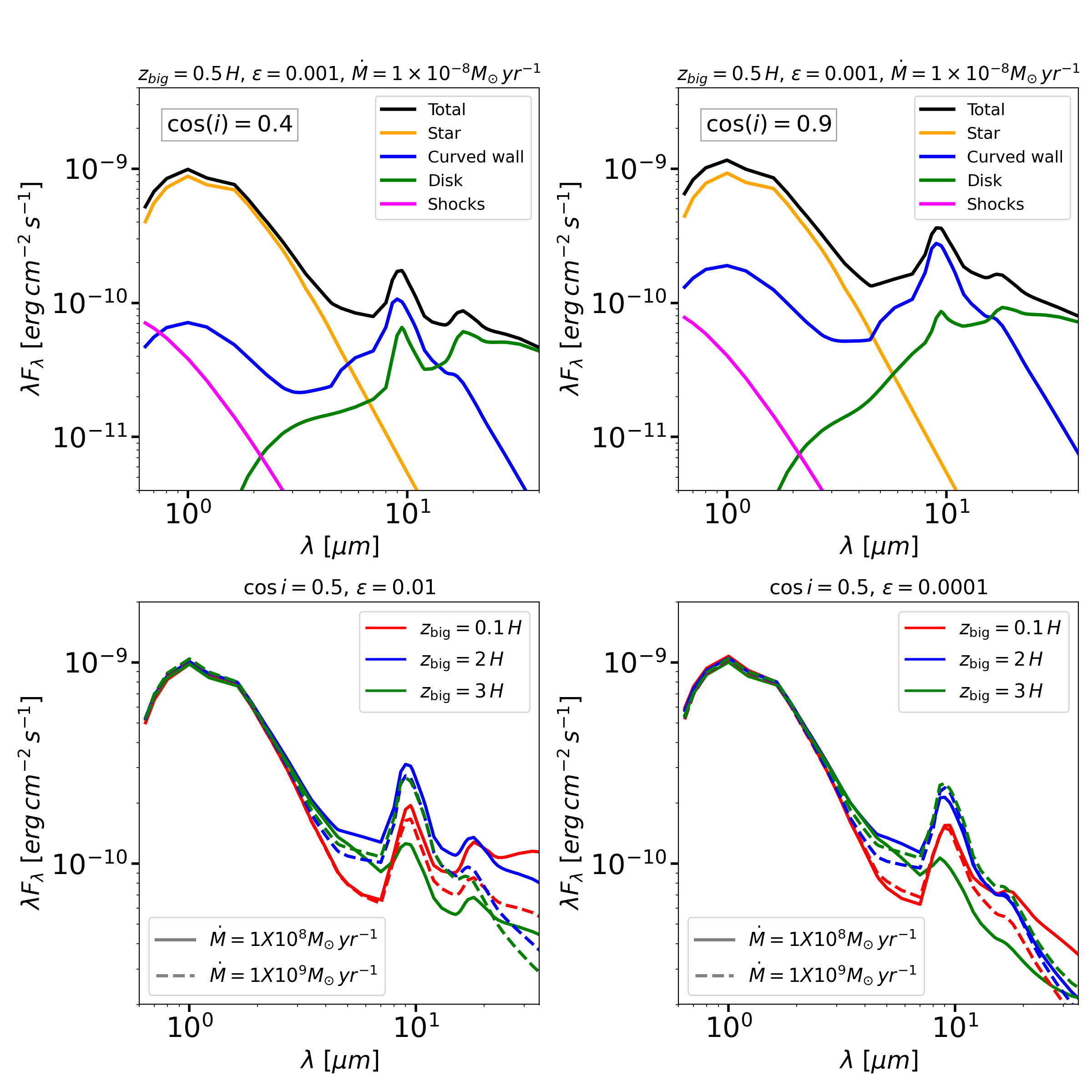}
    \caption{SEDs of disk models with curved walls. The upper panels are models with $\epsilon=0.001$, $\dot{M}=1 \times 10 ^{-8} M_{\odot} \, {\rm yr}^{-1}$, and $z_{\rm big}=0.5$ H. Two inclinations are shown: $\cos(i)=0.4$ ($i=66.4^{o}$, upper left) and  $\cos(i)=0.9$ ($i=25.8^{o}$, upper right). Both panels show the contribution from the star (yellow), the curved wall (blue), the disk (green), the accretion shocks (magenta), and the total  (black). The lower panels show models with $\cos(i)=0.5$, and two values of $\epsilon$: 0.01 (lower left) and 0.0001 (lower right). The colors indicate $z_{\rm big}$ values: 0.1 (red), 2 (blue), and 3 H (green), and the line style corresponds to $\dot{M}=1 \times 10 ^{-8} M_{\odot} \, {\rm yr}^{-1}$ (solid) and $\dot{M}=1 \times 10 ^{-9} M_{\odot} \, {\rm yr}^{-1}$ (dashed).}
    \label{fig_seds_panels}
\end{figure*}

To illustrate this, Figure \ref{fig_seds_panels} shows the spectral energy distributions (SEDs) of disk models with curved walls- In the upper panels, we show the SEDs for two inclinations: $\cos (i)=0.4$ (left) and $\cos (i)=0.9$ (right), together with the contributions from the star, disk, shocks, the curved wall, and the total emission. These panels correspond to models with $\dot{M}$= $1 \times 10 ^{-8} M_{\odot} \, {\rm yr}^{-1}$, $\epsilon=0.001$ and $z_{\rm big}=0.5$ H. We find that the emission from the curved wall is substantially higher for nearly face-on disks ($\cos (i)=0.9$ or $i=25.8^{o}$) than for more inclined systems ($\cos (i) =0.4$ or $i=66.4^{o}$). This is due to two effects: 1) at low inclinations, a larger fraction of the curved wall surface is exposed to the observer, i.e., there is little to no self-ocultation of wall elements by other parts of the wall; and 2) there is no disk material along the line of sight that can absorb the emerging radiation from the wall. At higher inclinations, part of the wall becomes self-hidden, and the larger column of disk material attenuates the wall emission. Note that our calculation of the curved wall emission (blue line) includes stellar scattered light off the wall surface, visible as a bump at stellar wavelengths. The contribution of this component depends on inclination: it accounts for $\sim 7 \%$ of the total near-IR flux for the model with $\cos(i)=0.4$, and $\sim 20 \%$ for the model with $\cos(i)=0.9$, for the same reasons discussed above. This near-IR excess produced by the wall is also seen in Figure \ref{fig_walls_SEDS} where we plotted the star and the wall's emission.

The lower panels of Figure \ref{fig_seds_panels} show the SEDs for models with different values of $z_{\rm big}$, indicated by different colors. These panels correspond to models with $\cos(i)=0.5$, $\epsilon=0.01$ (lower left) and $\epsilon=0.0001$ (lower right). In both panels, the line style indicates the adopted mass accretion rate: $\dot{M}=1\times10^{-8} M_{\odot} \,\mathrm{yr^{-1}}$ (solid lines) and $\dot{M}=1\times10^{-9} M_{\odot}\, \mathrm{yr^{-1}}$ (dashed lines). As discussed above, increasing $z_{\rm big}$ generally enhances the wall emission; however, an exception is seen for $z_{\rm big}=3$ H, where the opposite effect is observed in models with $\dot{M}$= $1 \times 10 ^{-8} M_{\odot} \, {\rm yr}^{-1}$. In addition, increased dust settling leads to a decrease in both the wall and the disk emission, with this effect becoming more pronounced at longer wavelengths ($\lambda \gtrsim 10 \, \mu {\rm m}$). From Figure \ref{fig_seds_panels} we see that the differences between models become more pronounced at wavelengths $\gtrsim 3 \, \mu{\rm m}$. Therefore, mid-infrared observations (e.g. JWST/MIRI, Spitzer/IRS), as well as IRAC, MIPS 24, or WISE photometry, provide strong diagnostic power to distinguish between models, and to perform detailed SED fitting. In particular, the shape and amplitude of the mid-IR excess are sensitive to the wall geometry, dust grain sizes, composition, and vertical distribution of grains, especially in the innermost disk regions. These diagnostics are expected to break degeneracies that remain in the near-IR regime. This will be explored in future work.\\

\section{Discussion}
\label{sec_discussion}

\subsection{Comparison with observations: near-IR colors of CTTSs}
\label{subsec_comparison}

\begin{figure*}
    \centering
    \includegraphics[scale=0.57]{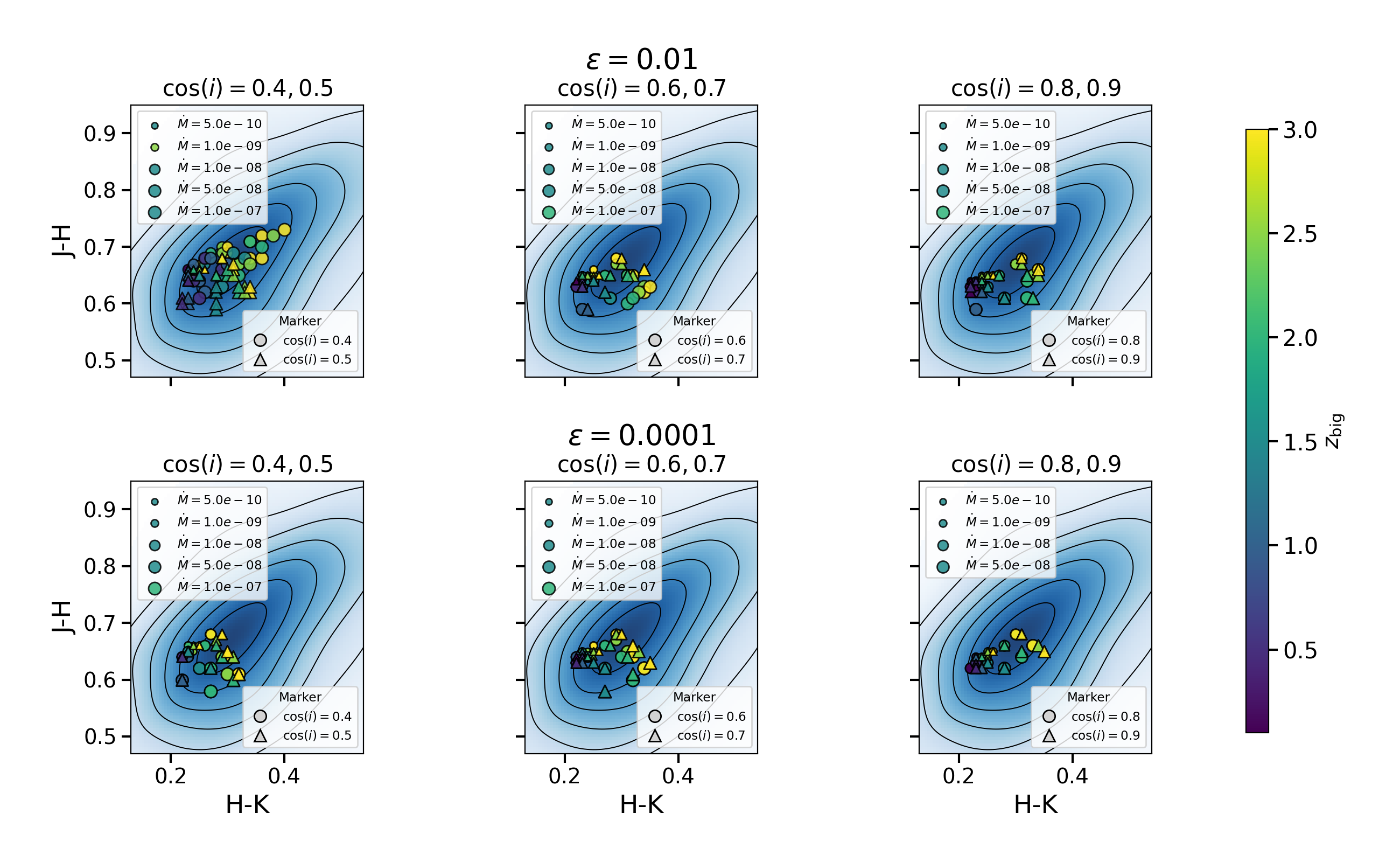} 
    \caption{KDE maps of the observed dereddened JHK colors (blue region), with relative density isocontours shown at values of 9, 7.5, 6, 4.5, 3, and 1.5,
    from innermost (highest relative density) to outermost (lowest relative density). The colored symbols correspond to   
    models with curved walls for different $\cos(i)$ as indicated by the labels. The first row shows models with $\epsilon=0.01$ and the second row models with $\epsilon=0.0001$. The size of the symbol increases with $\dot{M}$  and the color indicates the values for $z_{\rm big}$. We show models for a $0.5 \, M_{\odot}$, 1 Myr old star.}
    \label{fig_colores_modelos}
\end{figure*}

To test our models against the observations, particularly at near-IR wavelengths where the wall dominates the emission (Figure \ref{fig_seds_panels}), we compared the dereddened 2MASS JHK colors of the CTTS sample (Figure~\ref{fig:KDE_obs}) with synthetic colors from the disk-model grid described in Section~\ref{subsec_Dalessio_models}, which includes emission from the curved wall. Synthetic JHK magnitudes were computed by convolving the model fluxes with the 2MASS J, H, and K transmission curves, and colors were obtained from these magnitudes. Intrinsic photospheric colors were adopted from \citet{Kenyon_1995}.\\ 
In Figure \ref{fig_colores_modelos}, we show the distribution of synthetic colors in the ${\rm [J-H]}$ vs ${\rm[H-K]}$ plane for disk models with $\epsilon=0.01$ (upper panels) and $\epsilon=0.0001$ (lower panels), overlaid on the KDE map of the CTTS sample (Figure~\ref{fig:KDE_obs}). The black lines are KDE isocontours, i.e., lines of constant KDE value, drawn at six levels automatically selected by \texttt{matplotlib.contour} from the range of KDE values within the displayed domain. From the innermost to the outermost contour, the corresponding KDE values are 9, 7.5, 6, 4.5, 3, and 1.5. These contours do not represent enclosed-probability regions, but are used to visualize the shape of the CTTS color-color distribution. The KDE provides a smoothed estimate of the observational density in the color–color diagram. Therefore, models lying within high-density isocontours trace regions of higher observational concentration in color–color space.

Additional $\epsilon$ values within the parameter space described above were also considered when comparing the models with the observations; however, only two values are shown in Figure \ref{fig_colores_modelos} for reference. 
Each panel corresponds to different inclinations $\cos(i)$, the symbol size increases with increasing mass accretion rate $\dot{M}$, while the color scale indicates $z_{\rm big}$ values ranging from 0.1 H (blue) to 3 H (yellow). For intermediate inclinations $0.4 \leq \cos(i)\leq 0.5$, models with lower $\epsilon$ values (i.e., with stronger dust settling) systematically produce bluer colors, as expected \citep{Dalessio_2006}. For less inclined disks ($\cos(i)\geq 0.6$), the impact of $\epsilon$ becomes less relevant, as stellar emission is not extincted by disk material along the line of sight. In addition, for a given inclination, the reddest colors are produced by models with high $\dot{M}$ and large $z_{\rm big}$ values. 

\begin{figure*}
    \centering
    \includegraphics[scale=0.32]{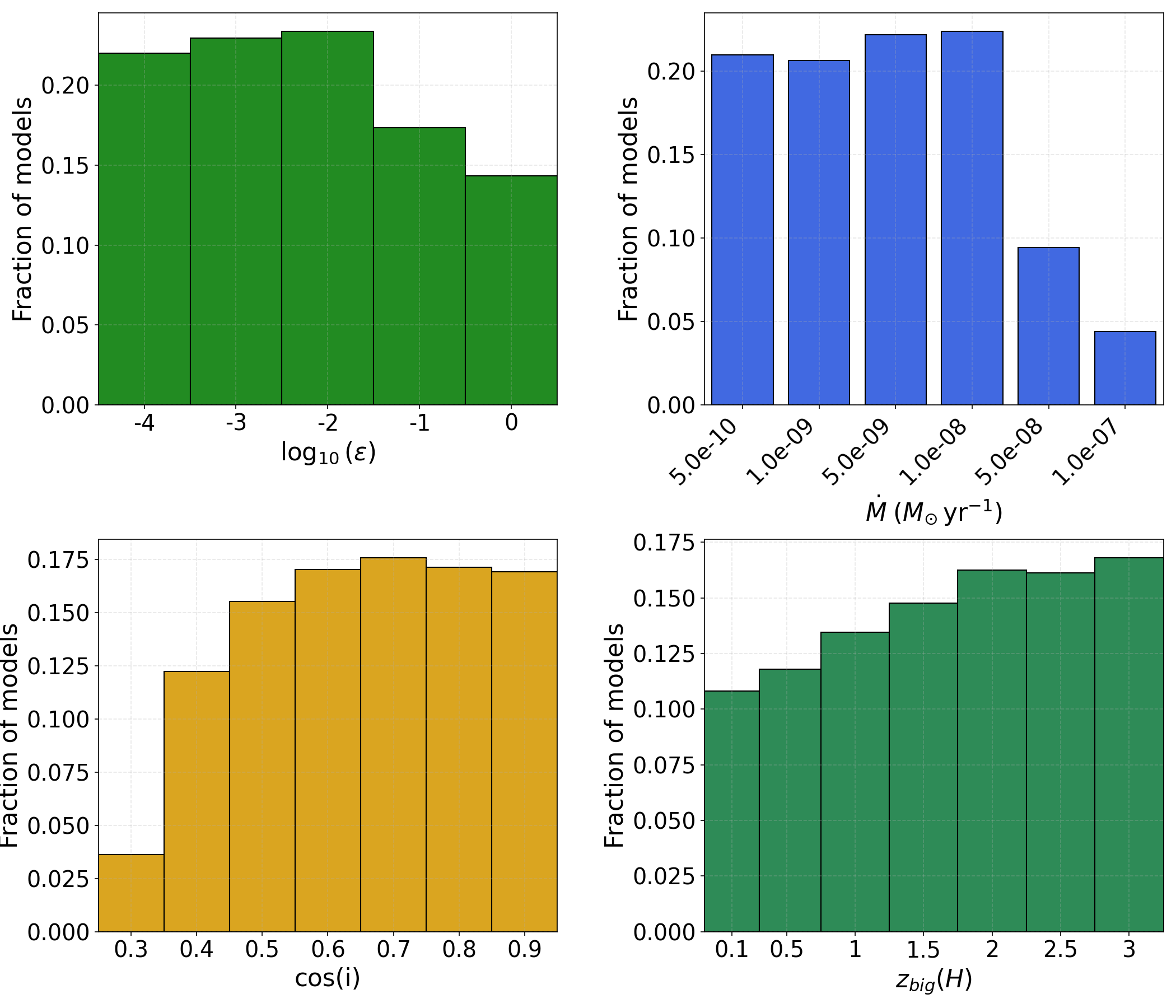} 
    \caption{Distributions of $\log(\epsilon)$ (upper left), $\dot{M}$ (upper right), $\cos(i)$ (lower left), and $z_{\rm big}$ (lower right), of the disk models that populate the region defined by the two innermost relative density isocontours (with KDE values 9 and 7.5) of the observations.}
    \label{fig_dist_param}
\end{figure*}

\begin{figure*}
    \centering
    \includegraphics[scale=0.32]{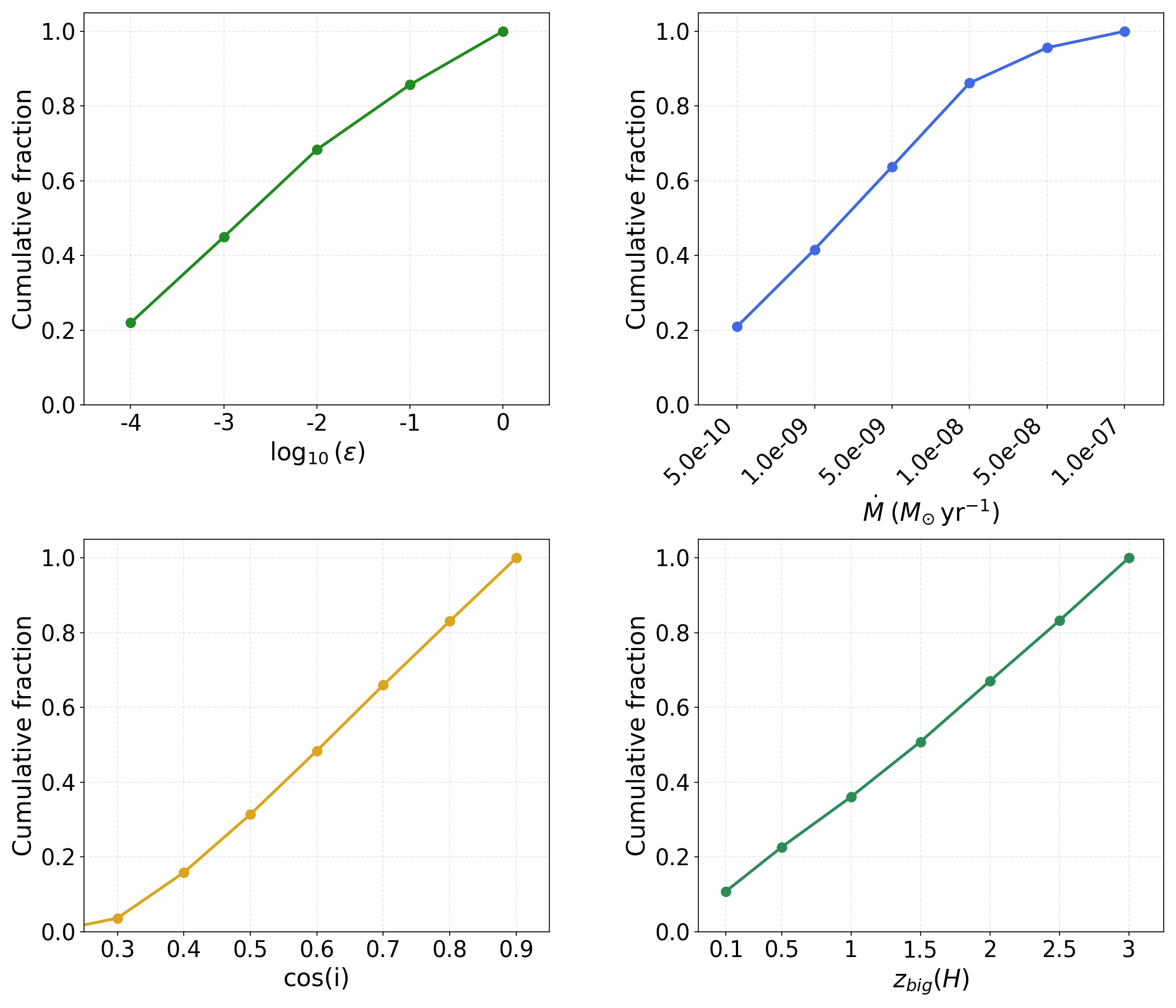} 
    \caption{Cumulative distributions of $\log(\epsilon)$ (upper left), $\dot{M}$ (upper right), $\cos(i)$ (lower left), and $z_{\rm big}$ (lower right), of the disk models that populate the region defined by the two innermost relative density isocontours (with KDE values 9 and 7.5) of the observations.}
    \label{fig_cum_dist_param2}
\end{figure*}

We selected the models that populate the region with the highest relative density of observations, i.e., inside the second innermost density iso-contour in the KDE map, corresponding to a KDE value of 7.5. A model was included in this subsample if its position in the ${\rm [J-H]}$ versus ${\rm [H-K]}$ plane lies within the KDE$<7.5$ contour of the observed sample. We then examined the distributions of the physical parameters for the selected models, shown in Figure \ref{fig_dist_param}, and the corresponding cumulative distributions in Figure \ref{fig_cum_dist_param2}. In the upper left panel, the distribution of $\epsilon$ values suggests there is a trend toward models with strong dust settling, i.e., with $\epsilon \leq 0.01$ ($68\%$ have $\epsilon \leq 0.01$ or $\log(\epsilon) \leq -2$). This is consistent with observational evidence for dust settling in protoplanetary disks, as opposed to models with no settling, $\epsilon = 1$ \citep[e.g.,][]{Manzo_2020}. The upper right panel shows that the distribution of mass accretion rates is approximately flat over the range $\dot{M} \in [5 \times 10^{-10}, 1 \times 10^{-8} ] \, M_{\odot} \, {\rm yr}^{-1}$, which is typical of accretion rates measured in $0.3 \, M_{\odot}-0.5 \ M_{\odot}$ T Tauri stars \citep{Manara_2023}. They represent $\sim 87\%$ of the models,  and the rest ($\sim 13\%$) have accretion rates $\dot{M} \geq 1 \times 10^{-8} \ M_{\odot}\,\mathrm{yr}^{-1}$. The distribution of $\cos(i)$, shown in the lower left panel, indicates that models with relatively low inclinations, $\cos(i) \geq 0.5$, or equivalently $i \leq 60^{\circ}$, dominate the selected sample ($63\%$ have $i \leq 60^{\circ}$). This is expected because disks observed at higher inclinations, $\cos(i) \leq 0.5$, or $i\geq 60^{\circ}$, tend to produce redder colors due to the larger column of disk material along the line of sight. The maximum inclination found in this subsample is $i = 72.5^{\circ}$, corresponding to $\cos(i) = 0.3$, since no models with higher inclinations, $\cos(i) < 0.3$, fall within the region defined by the second innermost KDE iso-contour (KDE$<7.5$). In contrast, the distribution of $z_{\mathrm{big}}$, shown in the lower right panel, reveals that the observations are consistent with a large fraction of disks having large values, $z_{\mathrm{big}} \geq 1$ H. In particular, approximately $78\%$ of the models require $z_{\mathrm{big}} \geq 0.5$ H, while about $30\%$ correspond to $z_{\mathrm{big}} \geq 2$ H, placing large grains close to the upper layers of the disk. This is relevant because $z_{\mathrm{big}}$ measures the vertical extent of the millimeter-sized grain layer in the disk wall. Thus, in a large fraction of disks, these large grains must be vertically extended up to heights of $1-3$ H.
Overall, our analysis indicates that large values of $z_{\mathrm{big}}$ are required to reproduce the observed near-IR colors. In particular, the models favor a population of large grains, with $a_{\max} \sim 1$ mm, extending to heights comparable to one or a few local disk scale heights. These heights correspond to approximately $0.8-2.5 \ R_{\ast}$, or $0.008-0.02$ au, at innermost disk radii of $\lesssim  0.2$ au. At the same time, most of the selected models require mass accretion rates and dust-settling values consistent with typical values measured in T Tauri stars.

\subsection{Curved walls to test turbulence}

The midplane in the innermost regions of protoplanetary disks is one of the most relevant sites in terms of solids processing and planet formation, and yet remains almost inaccessible to direct observations. 
The high densities make these regions optically thick, preventing radiation from the midplane from escaping.
In particular, determining the degree of turbulence in the midplane close to the star has been a challenge.
Current studies of turbulence have been made using molecular line observations from various species, but they originate in the disk's upper layers and at radii larger than the wall location. Recently, \citet{Paneque_2024} analyzed CN and ${\text C_{2}}$H emission in the disk around IM Lup, 
tracing regions at radii $\gtrsim 150$ au,
and found evidence of vertical gradients in turbulence, with enhanced turbulence in the disk upper layers. They suggested that the magnetorotational instability \citep[MRI,][]{Balbus_1991} may be active in the inner disk. Consistently, \citet{Flaherty_2024} found that IM Lup is a highly turbulent disk based on ALMA observations of multiple CO transitions, attributing these results to the MRI or hydrodynamical instabilities. Similarly, \cite{Long_2024} studied the disk around DM Tau using HCO$^{+}$, ${\text H^{13}}$CO$^{+}$, and ${\text N_{2}}$H$^{+}$ emission, and although they could not conclude that the disk turbulence is MRI driven, they reported relatively high turbulent velocities. 
However, other ALMA studies have reported disks with low levels of turbulence. For example \citet{Flaherty_2020} detected turbulence in DM Tau, but found no evidence of turbulence in MWC 480 and V4046 Sgr. 
Importantly, these studies probe disk regions at radii $\sim 20-50$ au or beyond, which are much farther out than the innermost disk and the wall.

The dust sublimation region, i.e., the inner wall, arguably offers the only way to probe the disk conditions down to the midplane in the innermost regions.
In this regard, Figures \ref{fig_walls_SEDS} and \ref{fig_seds_panels} show 
that the emission of the curved wall is highly sensitive to the value of $z_{\rm big}$, that is, to the height of the layer of millimeter-sized dust grains above the midplane. Furthermore, Figures \ref{fig_dist_param} and \ref{fig_cum_dist_param2} shows that models with $z_{\rm big}\geq0.5$ H and $z_{\rm big}\geq2$ H account for $\sim 78\%$ and $\sim 30\%$, respectively, of the models that populate the regions of highest relative density in the JHK color-color diagram (Figure \ref{fig_colores_modelos}). One possible explanation for this result is that there must be an active mechanism capable of stirring and lifting large millimeter-sized dust grains from the midplane at the wall location in disks with $z_{\rm big}>0.5$ H. Turbulence is one of the most likely candidates responsible for this effect. 
This result is particularly relevant in light of previous studies showing that disks are highly settled at ages as young as 1 Myr \citep[e.g.][]{Manzo_2020, Rilinger_2023}, based on SED fitting and analysis of IR colors. 
However, if turbulence operates in the innermost disk regions, its presence in the midplane may be traced by modeling the near-IR excess observed in T Tauri stars, which is primarily dominated by emission from the curved wall (Figure \ref{fig_seds_panels}).

Tracing turbulence in the inner disk has important implications, as turbulent mixing can regulate the formation and vertical transport of various molecules at different heights, into zones where planets are forming, likely near the midplane \citep{Ilgner_2006, Vaikundaraman_2025}.  This process might help explain why certain disks exhibit a more carbon-rich chemistry than others, as revealed by JWST spectra \citep{Kanwar_2024, Colmenares_2024, Kanwar_2024b}. Turbulence may also influence the efficiency of planetary embryo growth, the retention of planetary atmospheres, and ultimately the final architecture of planetary systems \citep{Bitsch_2015}. More broadly, turbulence may also play a central role in several processes, e.g.: (1) the transfer and redistribution of angular momentum throughout the disk; (2)  the growth and migration of solid particles, thereby affecting exoplanet formation; and (3) the efficiency of mass accretion onto the central star \citep{Armitage_2011, Turner_2014}.

The physical origin of turbulence in protoplanetary disks remains uncertain. While it may be MRI driven, other mechanisms have also been proposed, including magnetically driven winds \citep{Bai_2013} and the vertical shearing instability \citep{Lyra_2019}. If MRI-driven turbulence is present in the inner disk, it would naturally account for the vertical stirring of large grains responsible for the observed near-IR excess, without the need to invoke additional processes such as disk winds. 
Detailed studies of the inner disk emission in large samples of young stars are needed to better constrain turbulence at the midplane in the innermost regions,
and to discriminate between the different physical mechanisms proposed, thereby enabling stronger conclusions. Ultimately, probing the innermost disk regions offers a promising pathway to link the disk physical processes with the observed diversity of exoplanets.\\

\section{Summary and conclusions}
\label{sec_summary}

In this work, we present a new model for the geometry and emission of curved walls or inner rims in T Tauri disks. We use the D'Alessio models, which enforce hydrostatic equilibrium to find the disk structure in a self-consistent approach, to determine the wall properties. We explore how the emission and geometry depend on the different disk parameters, and test our disk models with curved walls using near-IR dereddened colors in the JHK bands of a large sample of CTTSs. Our main results are summarized here:

\begin{itemize}
    \item  For our fiducial model with a $0.5 \, M_{\odot}$ central star, we find that the curved wall spans radially from $\sim 0.12$ to $0.33$ au. These values, which are consistent with observational estimates of wall locations, vary depending on the specific parameters of the disk.
    \item The curvature of the inner rim is caused by vertical gradients in density, temperature, and opacity, which is determined by the relative abundances of small and large grains at different heights. Additionally, the sublimation temperature varies with local density.\\
    \item The emission from the curved wall dominates the near-IR of the SEDs and is highly sensitive to the parameter $z_{\rm big}$, which represents the height of the midplane layer containing large mm-sized grains. As $z_{\rm big}$ increases, the wall flux also increases; however, for $z_{\rm big} \gtrsim 3$ H, the effect is the opposite, and the $10\,\mu{\rm m}$ silicate band becomes less prominent. This occurs because the fraction of small, atmospheric grains, responsible for producing this feature, decreases.\\
    \item The models show that for disks with low inclinations ($\cos(i) \sim 0.9$), the emission from the curved wall is stronger compared to disks with higher inclinations ($\cos(i) \sim 0.4$). This is because, at inclinations close to face-on, the entire surface area of the wall, observed as a bright ring, is fully visible to the observer. Additionally, in these cases, the wall’s emission is unaffected by extinction from dust in the disk.\\
    \item We use near-IR colors of a large sample of CTTSs in Taurus, IC 348, and Orion  to test our models with curved walls. We are able to reproduce the distribution of dereddened colors in the [J-H] vs [H-K] diagram using synthetic colors from the models.
    \item We find that disk models with different combinations of mass accretion rate $\dot{M}$, dust settling $\epsilon$, and inclination $\cos(i)$ can populate the region in the JHK diagram with the highest relative density of observations. However, $78\%$ and $30\%$ of the models in that region have $z_{\rm big}\geq0.5$ H and $z_{\rm big}\geq 2$ H, respectively. This indicates that large grains are not entirely confined to the midplane in the innermost radii $\lesssim 0.2$ au.\\
    \item Previous studies have found that the millimeter-sized dust grains settle to the midplane,
    so a mechanism is needed for stirring and lifting the large mm dust grains up to 3 H in the disk's innermost regions ($\lesssim 0.2$ au). We propose that this mechanism is turbulence.\\
    \item Modeling of the near-IR excess of CTTSs, which is dominated by the wall, potentially allows us to trace the presence of turbulence in the disks' innermost regions, which has important implications in chemistry, planet formation, and the transfer of angular momentum.\\
\end{itemize}

\section*{Acknowledgements}
   We thank the anonymous referee for valuable comments and suggestions that improved the presentation of this paper. E.M.M. acknowledges the support from a postdoctoral fellowship from Secretaría de Ciencia, Humanidades, Tecnología e Innovación (SECIHTI, formerly CONAHCyT), and the University of Michigan for the facilities provided for the development of this work. This work was partially supported by grant NASA XRP 80NSSC2K0151. N.C. acknowledges partial support from grant NSF grant PHY-2309135 to the Kavli Institute for Theoretical Physics (KITP). This support provided a stay at KITP, which inspired ideas for this paper. J.H. acknowledges support from the UNAM-DGAPA-PAPIIT research projects IG-101723 and IN110126. We thank the Models and Observations of Disk Evolution in Latin America (MODELA) collaboration for their valuable discussions and contributions to this work.


\begin{thebibliography}{}

\bibitem[Akeson et al.(2005)]{Akeson_2005} Akeson, R.~L., Boden, A.~F., Monnier, J.~D., et al.\ 2005, \apj, 635, 2, 1173. doi:10.1086/497436
\bibitem[Anthonioz et al.(2015)]{Anthonioz_2015} Anthonioz, F., M{\'e}nard, F., Pinte, C., et al.\ 2015, \aap, 574, A41. doi:10.1051/0004-6361/201424520
\bibitem[Armitage(2011)]{Armitage_2011} Armitage, P.~J.\ 2011, \araa, 49, 1, 195. doi:10.1146/annurev-astro-081710-102521
\bibitem[Bai \& Stone(2013)]{Bai_2013} Bai, X.-N. \& Stone, J.~M.\ 2013, \apj, 769, 1, 76. doi:10.1088/0004-637X/769/1/76
\bibitem[Balbus \& Hawley(1991)]{Balbus_1991} Balbus, S.~A. \& Hawley, J.~F.\ 1991, \apj, 376, 214. doi:10.1086/170270
\bibitem[Birnstiel et al.(2015)]{Birnstiel_2015} Birnstiel, T., Andrews, S.~M., Pinilla, P., et al.\ 2015, \apjl, 813, 1, L14. doi:10.1088/2041-8205/813/1/L14
\bibitem[Birnstiel et al.(2011)]{Birnstiel_2011} Birnstiel, T., Ormel, C.~W., \& Dullemond, C.~P.\ 2011, \aap, 525, A11. doi:10.1051/0004-6361/201015228
\bibitem[Bitsch et al.(2015)]{Bitsch_2015} Bitsch, B., Johansen, A., Lambrechts, M., et al.\ 2015, \aap, 575, A28. doi:10.1051/0004-6361/201424964
\bibitem[Brice{\~n}o et al.(2019)]{Briceno_2019} Brice{\~n}o, C., Calvet, N., Hern{\'a}ndez, J., et al.\ 2019, \aj, 157, 85. doi:10.3847/1538-3881/aaf79b
\bibitem[Calvet \& Gullbring(1998)]{Calvet_1998} Calvet, N. \& Gullbring, E.\ 1998, \apj, 509, 2, 802. doi:10.1086/306527
\bibitem[Cardelli et al.(1989)]{Cardelli_1989} Cardelli, J.~A., Clayton, G.~C., \& Mathis, J.~S.\ 1989, \apj, 345, 245. doi:10.1086/167900
\bibitem[Cevallos Soto \& Zhu(2025)]{Cevallos_2025} Cevallos Soto, A. \& Zhu, Z.\ 2025, , arXiv:2505.03701. doi:10.48550/arXiv.2505.03701
\bibitem[Chrenko et al.(2024)]{Chrenko_2024} Chrenko, O., Flock, M., Ueda, T., et al.\ 2024, \aj, 167, 3, 124. doi:10.3847/1538-3881/ad234d
\bibitem[Colmenares et al.(2024)]{Colmenares_2024} Colmenares, M.~J., Bergin, E.~A., Salyk, C., et al.\ 2024, \apj, 977, 2, 173. doi:10.3847/1538-4357/ad8b4f
\bibitem[D'Alessio et al. (1998)]{Dalessio_1998} D'Alessio, P.; Cant\'o, J.; Calvet, N.; Lizano, S. \ 1998, \apj, 500, 411
\bibitem[D'Alessio et al. (1999)]{Dalessio_1999} 	
D'Alessio, P.; Calvet, N.; Hartmann, L.; Lizano, S.; Cant\'o, J., \ 1999, \apj, 527, 893
\bibitem[D'Alessio et al. (2001)]{Dalessio_2001} 	
D'Alessio, P.; Calvet, N.; Hartmann, L. \ 2001, \apj, 553, 321
\bibitem[D'Alessio et al.(2004)]{Dalessio_2004} D'Alessio, P., Calvet, N., Hartmann, L., et al.\ 2004, Star Formation at High Angular Resolution, 221, 403. doi:10.48550/arXiv.astro-ph/0309590
\bibitem[D'Alessio et al. (2005)]{Dalessio_2005} D'Alessio, P.; Hartmann, L.; Calvet, N.; Franco-Hernández, R.; et al. \ 2005, \apj, 621, 461
\bibitem[D'Alessio et al. (2006)]{Dalessio_2006}	
D'Alessio, P.; Calvet, N.; Hartmann, L.; Franco-Hernández, R.; Servín, H. \ 2006, \apj, 638, 314
\bibitem[Dullemond \& Monnier(2010)]{Dullemond_2010} Dullemond, C.~P. \& Monnier, J.~D.\ 2010, \araa, 48, 205. doi:10.1146/annurev-astro-081309-130932
\bibitem[Espaillat et al.(2014)]{Espaillat_2014} Espaillat, C., Muzerolle, J., Najita, J., et al.\ 2014, Protostars and Planets VI, 497. doi:10.2458/azu\_uapress\_9780816531240-ch022
\bibitem[Esplin \& Luhman(2019)]{Esplin_2019} Esplin, T.~L. \& Luhman, K.~L.\ 2019, \aj, 158, 54. doi:10.3847/1538-3881/ab2594
\bibitem[Flaherty et al.(2020)]{Flaherty_2020} Flaherty, K., Hughes, A.~M., Simon, J.~B., et al.\ 2020, \apj, 895, 2, 109. doi:10.3847/1538-4357/ab8cc5
\bibitem[Flaherty et al.(2024)]{Flaherty_2024} Flaherty, K., Hughes, A.~M., Simon, J.~B., et al.\ 2024, \mnras, 532, 1, 363. doi:10.1093/mnras/stae1480
\bibitem[Flock et al.(2016)]{Flock_2016} Flock, M., Fromang, S., Turner, N.~J., et al.\ 2016, \apj, 827, 2, 144. doi:10.3847/0004-637X/827/2/144
\bibitem[Flock et al.(2017)]{Flock_2017} Flock, M., Fromang, S., Turner, N.~J., et al.\ 2017, \apj, 835, 2, 230. doi:10.3847/1538-4357/835/2/230
\bibitem[Flock et al.(2019)]{Flock_2019} Flock, M., Turner, N.~J., Mulders, G.~D., et al.\ 2019, \aap, 630, A147. 
\bibitem[Flock et al.(2025)]{Flock_2025} Flock, M., Chrenko, O., Ueda, T., et al.\ 2025, \aap, 701, A259. doi:10.1051/0004-6361/202453124
doi:10.1051/0004-6361/201935806
\bibitem[Furlan et al.(2011)]{Furlan_2011} Furlan, E., Luhman, K.~L., Espaillat, C., et al.\ 2011, \apjs, 195, 1, 3. doi:10.1088/0067-0049/195/1/3
\bibitem[Galli et al.(2019)]{Galli_2019} Galli, P.~A.~B., Loinard, L., Bouy, H., et al.\ 2019, \aap, 630, A137. doi:10.1051/0004-6361/201935928
\bibitem[Gole et al.(2020)]{Gole_2020} Gole, D.~A., Simon, J.~B., Li, R., et al.\ 2020, \apj, 904, 2, 132. doi:10.3847/1538-4357/abc334
\bibitem[GRAVITY Collaboration et al.(2021)]{Gravity_2021} GRAVITY Collaboration, Perraut, K., Labadie, L., et al.\ 2021, \aap, 655, A73. 
doi:10.1051/0004-6361/202141624
\bibitem[Herbst(2008)]{Herbst_2008} Herbst, W.\ 2008, Handbook of Star Forming Regions, Volume I, 4, 372. 
\bibitem[Hern{\'a}ndez et al.(2007a)]{Hernandez_2007a} Hern{\'a}ndez, J., Hartmann, L., Megeath, T., et al.\ 2007, \apj, 662, 2, 1067. doi:10.1086/513735
\bibitem[Hern{\'a}ndez et al.(2007b)]{Hernandez_2007b} Hern{\'a}ndez, J., Calvet, N., Brice{\~n}o, C., et al.\ 2007, \apj, 671, 2, 1784. doi:10.1086/522882
\bibitem[Ibrahim et al.(2023)]{Ibrahim_2023} Ibrahim, N., Monnier, J.~D., Kraus, S., et al.\ 2023, \apj, 947, 2, 68. doi:10.3847/1538-4357/acb4ea
\bibitem[Ilgner \& Nelson(2006)]{Ilgner_2006} Ilgner, M. \& Nelson, R.~P.\ 2006, \aap, 445, 1, 223. doi:10.1051/0004-6361:20053867
\bibitem[Kanwar et al.(2024a)]{Kanwar_2024} Kanwar, J., Kamp, I., Woitke, P., et al.\ 2024, \aap, 681, A22. doi:10.1051/0004-6361/202346262
\bibitem[Kanwar et al.(2024b)]{Kanwar_2024b} Kanwar, J., Kamp, I., Jang, H., et al.\ 2024, \aap, 689, A231. doi:10.1051/0004-6361/202450078
\bibitem[Kenyon \& Hartmann(1995)]{Kenyon_1995} Kenyon, S.~J. \& Hartmann, L.\ 1995, \apjs, 101, 117. doi:10.1086/192235
\bibitem[Kluska et al.(2020)]{Kluska_2020} Kluska, J., Berger, J.-P., Malbet, F., et al.\ 2020, \aap, 636, A116. doi:10.1051/0004-6361/201833774
\bibitem[Lada et al.(2006)]{Lada_2006} Lada, C.~J., Muench, A.~A., Luhman, K.~L., et al.\ 2006, \aj, 131, 1574. doi:10.1086/499808
\bibitem[Luhman et al.(2016)]{Luhman_2016} Luhman, K.~L., Esplin, T.~L., \& Loutrel, N.~P.\ 2016, \apj, 827, 52. doi:10.3847/0004-637X/827/1/52
\bibitem[Luhman(2023)]{Luhman2023} Luhman, K.~L.\ 2023, \aj, 165, 2, 37. doi:10.3847/1538-3881/ac9da3
\bibitem[Lyra \& Umurhan(2019)]{Lyra_2019} Lyra, W. \& Umurhan, O.~M.\ 2019, \pasp, 131, 1001, 072001. doi:10.1088/1538-3873/aaf5ff
\bibitem[Isella \& Natta(2005)]{Isella_2005} Isella, A. \& Natta, A.\ 2005, \aap, 438, 899. doi:10.1051/0004-6361:20052773
\bibitem[Kama et al.(2009)]{Kama_2009} Kama, M., Min, M., \& Dominik, C.\ 2009, \aap, 506, 3, 1199. doi:10.1051/0004-6361/200912068
\bibitem[Kounkel et al.(2018)]{Kounkel_2018} Kounkel, M., Covey, K., Su{\'a}rez, G., et al.\ 2018, \aj, 156, 3, 84. doi:10.3847/1538-3881/aad1f1
\bibitem[Lazareff et al.(2017)]{Lazareff_2017} Lazareff, B., Berger, J.-P., Kluska, J., et al.\ 2017, \aap, 599, A85. doi:10.1051/0004-6361/201629305
\bibitem[Long et al.(2024)]{Long_2024} Long, D.~E., Cleeves, L.~I., Adams, F.~C., et al.\ 2024, \apj, 972, 1, 88. doi:10.3847/1538-4357/ad5c67
\bibitem[Manara et al.(2023)]{Manara_2023} Manara, C.~F., Ansdell, M., Rosotti, G.~P., et al.\ 2023, Protostars and Planets VII, 534, 539. doi:10.48550/arXiv.2203.09930

\bibitem[Manzo-Mart{\'\i}nez et al.(2020)]{Manzo_2020} Manzo-Mart{\'\i}nez, E., Calvet, N., Hern{\'a}ndez, J., et al.\ 2020, \apj, 893, 56. doi:10.3847/1538-4357/ab7ead
\bibitem[McClure et al.(2013)]{McClure_2013} McClure, M.~K., D'Alessio, P., Calvet, N., et al.\ 2013, \apj, 775, 114. doi:10.1088/0004-637X/775/2/114
\bibitem[Mendigut{\'\i}a et al.(2024)]{Mendigutia_2024} Mendigut{\'\i}a, I., Lillo-Box, J., Vioque, M., et al.\ 2024, \aap, 686, L1. doi:10.1051/0004-6361/202449368
\bibitem[Mori et al.(2025)]{Mori_2025} Mori, S., Kunitomo, M., \& Ogihara, M.\ 2025, \aap, 697, A192. doi:10.1051/0004-6361/202453362
\bibitem[Natta et al.(2001)]{Natta_2001} Natta, A., Prusti, T., Neri, R., et al.\ 2001, \aap, 371, 186. doi:10.1051/0004-6361:20010334
\bibitem[Ortiz-Le{\'o}n et al.(2018)]{Ortiz_2018} Ortiz-Le{\'o}n, G.~N., Loinard, L., Dzib, S.~A., et al.\ 2018, \apj, 865, 1, 73. doi:10.3847/1538-4357/aada49
\bibitem[Paneque-Carre{\~n}o et al.(2024)]{Paneque_2024} Paneque-Carre{\~n}o, T., Izquierdo, A.~F., Teague, R., et al.\ 2024, \aap, 684, A174. doi:10.1051/0004-6361/202347757
\bibitem[Pollack et al.(1994)]{Pollack_1994} Pollack, J.~B., Hollenbach, D., Beckwith, S., et al.\ 1994, \apj, 421, 615. doi:10.1086/173677
\bibitem[Ribas et al.(2020)]{Ribas_2020} Ribas, {\'A}., Espaillat, C.~C., Mac{\'\i}as, E., et al.\ 2020, \aap, 642, A171. doi:10.1051/0004-6361/202038352
\bibitem[Rilinger et al.(2023)]{Rilinger_2023} Rilinger, A.~M., Espaillat, C.~C., Xin, Z., et al.\ 2023, \apj, 944, 1, 66. doi:10.3847/1538-4357/aca905
\bibitem[Shakura \& Sunyaev(1973)]{Shakura_1973} Shakura, N.~I. \& Sunyaev, R.~A.\ 1973, \aap, 24, 337. 
\bibitem[Skrutskie et al.(2006)]{2MASS_2006} Skrutskie, M.~F., Cutri, R.~M., Stiening, R., et al.\ 2006, \aj, 131, 2, 1163. doi:10.1086/498708
\bibitem[Sun et al.(2025)]{Sun_2025} Sun, M.-F., Xie, J.-W., Zhou, J.-L., et al.\ 2025, \aap, 699, A333. doi:10.1051/0004-6361/202553671
\bibitem[Tannirkulam et al.(2007)]{Tannirkulam_2007} Tannirkulam, A., Harries, T.~J., \& Monnier, J.~D.\ 2007, \apj, 661, 1, 374. doi:10.1086/513265
\bibitem[Turner et al.(2014)]{Turner_2014} Turner, N.~J., Fromang, S., Gammie, C., et al.\ 2014, Protostars and Planets VI, 411. doi:10.2458/azu\_uapress\_9780816531240-ch018
\bibitem[Ueda et al.(2017)]{Ueda_2017} Ueda, T., Okuzumi, S., \& Flock, M.\ 2017, \apj, 843, 1, 49. doi:10.3847/1538-4357/aa74b5
\bibitem[Ueda et al.(2019)]{Ueda_2019} Ueda, T., Flock, M., \& Okuzumi, S.\ 2019, \apj, 871, 1, 10. doi:10.3847/1538-4357/aaf3a1
\bibitem[Vaikundaraman et al.(2025)]{Vaikundaraman_2025} Vaikundaraman, V., Dr{k{a}}{\.z}kowska, J., Binkert, F., et al.\ 2025, \aap, 696, A215. doi:10.1051/0004-6361/202553850




\end{thebibliography}
\end{document}